\documentclass[preprint,amsmath,amssymb,aps,prd]{revtex4-2}
\usepackage{graphicx}
\usepackage{dcolumn}
\usepackage{bm}
\usepackage{amsmath}
\usepackage{hyperref}
\usepackage{hyperref}
\usepackage{multirow}
\usepackage{subcaption} 
\usepackage{cleveref}
\usepackage{caption} 
\usepackage{changes} 
\numberwithin{equation}{section} 

\begin{document}

\title{Ringdown Signatures of Dehnen Dark Matter Halos: Fluid Modes and Detectability with Space-Based Detectors}

\author{Manjia Liang$^{1, 2}$}
\author{Minghui Du$^{1}$}
\author{Qing Diao$^{1}$}
\author{Bo Liang$^{1}$}
\author{Ziren Luo$^{1}$}

\author{Peng Xu$^{1, 2, 3, 4}$}\email[Email: ]{pengxu@imech.ac.cn}
\author{Wei-Liang Qian$^{5, 6, 7}$}\email[Email: ]{wlqian@usp.br}
\author{Massimo Tinto$^{8}$}\email[Email: ]{massimo.tinto@gmail.com}

\affiliation{$^{1 }$Center for Gravitational Wave Experiment, National Microgravity Laboratory, Institute of Mechanics,
Chinese Academy of Sciences, Beijing 100190, China}
\affiliation{$^{2}$University of Chinese Academy of Sciences, Beijing 101400, China}

\affiliation{$^{3}$School of Fundamental Physics and Mathematical Sciences, Hangzhou Institute for Advanced Study, UCAS, Hangzhou 310024, China}

\affiliation{$^{4}$Lanzhou Center of Theoretical Physics, Lanzhou University, Lanzhou 730000, China}

\affiliation{$^{5}$Escola de Engenharia de Lorena, Universidade de S\~ao Paulo, 12602-810, Lorena, SP, Brazil}
\affiliation{$^{6}$Faculdade de Engenharia de Guaratinguet\'a, Universidade Estadual Paulista, 12516-410, Guaratinguet\'a, SP, Brazil}
\affiliation{$^{7}$ School of Physics and Technology, Nantong University, Nantong 226019, China}
\affiliation{$^{8}$Divisão de Astrofísica, Instituto Nacional de Pesquisas Espaciais, S. J. Campos 12227-010, SP, Brazil}

\date{\today}

\begin{abstract}
In this work, we investigate the feasibility of using ringdown waveforms from supermassive black holes immersed in dark-matter halos to extract both the intrinsic black-hole parameters and those characterizing the surrounding matter distribution with future space-based gravitational-wave detectors.
Building on the fully relativistic framework developed by Cardoso {\it et al.}, in which the dark-matter degrees of freedom are explicitly accounted for by minimal coupling to the gravitational sector, we construct numerical waveforms for a variety of Dehnen-type dark-matter profiles.
We then convert these simulated waveforms into realistic data streams for future space-based gravitational-wave observatories, consistently implementing the second-generation Time-Delay Interferometry scheme in the analysis.
We calculate the signal-to-noise ratios and perform a Bayesian parameter estimation to infer the model parameters, quantifying their measurability through the resulting posterior distributions.
Our results indicate that the presence of dark matter can induce sizable modifications to the waveforms through the appearance of fluid modes at late times.
Furthermore, dark-matter profiles with more pronounced spikes leave stronger imprints on the gravitational-wave signal, thereby enhancing the prospects for parameter inference with future space-based detectors such as LISA, Taiji, and TianQin.
\end{abstract}

\maketitle
\newpage

\section{Introduction}\label{sec:intro}

Dark matter~\cite{bertone2018history, arcadi2018waning, audren2014strongest} constitutes one of the most intriguing yet unresolved features of the standard cosmological model.
Despite robust observational evidence for a non-baryonic component interacting via gravity, its underlying particle-physics properties remain obscure.
Possible theoretical candidates include proposals for cold, warm, and ultra-light dark matter.
However, the absence of electromagnetic signatures makes dark matter extremely challenging to detect empirically. 
Consequently, developing robust and practical detection strategies is essential to discriminate among competing theoretical proposals based on experimental data.

With the rapid progress of gravitational-wave (GW) astronomy in recent years, probing dark matter directly with gravitational radiation has become a critical research area~\cite{abbott2016observation,PhysRevX.9.031040, cardoso2022gravitational, zhang2024detecting, eda2013new}.
Ongoing projects for space-based GW detectors, a cornerstone of future GW astronomy, primarily target astrophysical processes involving massive compact objects, such as supermassive black holes (SMBHs)~\cite{cai2024networks,taiji_0,luo2025progress,amaro2017laser}.
The LISA and Taiji planned missions are expected to detect GWs from black hole mergers with masses in the $10^{5}-10^7 M_{\odot}$ range, delivering signal-to-noise ratios (SNRs) of $10^2-10^4$.
This presents a compelling opportunity to extract the underlying information on dark matter from these exceptionally {\it loud} sources.
This endeavor, known as black hole spectroscopy, has been extensively explored and developed in recent years~\cite{agr-BH-spectroscopy-review-04}.

The quasinormal modes (QNMs) excited during the black hole ringdown phase~\cite{vishveshwara1970stability,nollert1999quasinormal,ching1995wave,berti2009quasinormal} offer a promising observational window to investigate the imprints of ambient dark matter on GW signals~\cite{berti2006gravitational,shi2024detectability,zhang2021parameter,barausse2014can, PhysRevD.100.044036}.  
Although the GW amplitude is significantly suppressed in this final stage of coalescence, this epoch is crucial for several reasons~\cite{cardoso2022black,spieksma2025black,spieksma2025black,speri2023measuring,leong2023detectability,yunes2011imprint}.  
First, the close-limit approximation reduces the two-body system to a single perturbed black hole~\cite{PhysRevD.60.104021,PhysRevLett.72.3297, agr-BH-spectroscopy-61}.  
Since QNMs are precisely determined by the metric of an isolated compact object, any deviation from these QNMs might provide a precise measure of the spacetime deformation induced by the dark matter environment.  
Moreover, recent analyses indicate that spectral instability~\cite{spectral-instability-review-20} leads to sizable deformations across the QNM spectrum~\cite{agr-qnm-instability-02, agr-qnm-instability-07, agr-qnm-instability-11, agr-qnm-Poschl-Teller-16, agr-qnm-Poschl-Teller-17, agr-qnm-instability-14, agr-qnm-instability-43, agr-qnm-instability-57, agr-qnm-instability-70, agr-qnm-instability-83, agr-qnm-instability-84, agr-qnm-echoes-50, agr-qnm-continued-fraction-40, agr-qnm-Poschl-Teller-30}, affecting even the low-lying modes~\cite{agr-qnm-instability-15, agr-qnm-instability-58, PhysRevD.110.084018, agr-qnm-instability-55, agr-qnm-instability-13} or producing a distinct branch of QNMs~\cite{agr-qnm-echoes-20, agr-qnm-echoes-45, agr-qnm-instability-65}.  
Furthermore, the coupling between the matter and gravitational sectors gives rise to $w$-modes and $f$-modes (also known as fluid modes)~\cite{cardoso2022gravitational}, originally discussed in the context of stellar QNMs~\cite{agr-qnm-star-10, agr-qnm-star-20}.
This suggests that the resulting frequency-domain deformation may non-trivially affect the time-domain GW signal.

The density distributions of dark matter halos surrounding SMBHs have been extensively studied in the literature~\cite{dehnen1993family, gondolo1999dark,navarro1997universal,sadeghian2013dark,zhao1996analytical}.
Many theoretical density profiles for dark-matter halos exist, most of them either exhibiting a central spike~\cite{navarro1997universal} or a flat core~\cite{burkert1995structure}. 
Phenomenologically, the Dehnen family~\cite{dehnen1993family} offers greater flexibility than other canonical profiles.
It can be controlled by a single parameter, $\gamma$, which allows the same functional form to describe very different dark-matter distributions, and subsequently facilitates the analysis across different models. 

Building on earlier studies rooted in Newtonian gravity with relativistic corrections~\cite{PhysRevD.75.024005,PhysRevD.102.064041}, Cardoso $\textit{et al.}$, proposed a fully relativistic framework and derived a spherically symmetric, static black hole metric embedded in a dark matter halo described by the Hernquist profile.  
Subsequently, by deriving the master equation for metric and matter perturbations, this model has been used to compute ring-down waveforms.  
In this framework, the dark matter distribution is anisotropic, modeled as a fluid satisfying a barotropic equation of state, with its degrees of freedom minimally coupled to the gravitational sector~\cite{kojima1992equations,allen1998gravitational,cardoso2022gravitational}.  
In particular, while the axial gravitational sector is effectively decoupled from the fluid, the polar sector is intrinsically coupled with the matter degrees of freedom.  
For the latter, the temporal evolution of initial perturbations is governed by a system of wavelike equations.  
As a result, the presence of dark matter can lead to more pronounced observational effects, manifesting as more persistent fluid modes in the resulting gravitational radiation.  
This theoretical setup has been used to investigate extreme mass-ratio inspiral (EMRI) waveforms~\cite{cardoso2022gravitational, agr-EMRI-51}, QNMs~\cite{agr-dark-matter-82}, echoes~\cite{liu2024probing}, and superradiance~\cite{agr-dark-matter-81}.  
However, quantitative constraints on the properties of the underlying SMBH and its dark matter halo, based on realistic ringdown GW signals, remain largely unexplored in the current literature.

From an empirical perspective, extracting the underlying properties of these gravitational systems also faces various challenges~\cite{Baghi:2022ucj}.
It demands not only high-precision instrument performance but also sophisticated data analysis techniques, for which accurate modeling plays a crucial role.
To address this,  the Time-Delay Interferometry (TDI) scheme~\cite{agr-TDI-review-02, armstrong1999time, agr-TDI-Wang-09, agr-TDI-Wang-22, agr-TDI-50} was proposed to reduce laser frequency noise in space-based GW detection.
The laser frequency noise, as the primary source of noise, is typically seven or eight orders of magnitude above the GW signals.
The TDI scheme cancels laser frequency noise by algebraically combining time-shifted raw data streams, effectively generalizing the principle of equal-arm interferometers.
Subsequently, the resulting data streams are synthesized as linear combinations of the GW signals together with the residual noises, referred to as the noise floor.
For the estimation of GW source parameters from noisy data, matched filtering based on the Bayesian statistical framework has been the most well-established technique since the first GW detection~\cite{Allen:2005fk,LIGOScientific:2016aoc}.
However, in the literature, many studies have considered relatively simplified models as the sources of GWs.
On the other hand, for more sophisticated gravitational systems, most analyses deal with the raw GW signals, without accounting for the TDI combinations, which inevitably complicate the scenario by time-shifting and folding the GW signals before they are processed.

The present study addresses these gaps. We investigate the feasibility of extracting metric parameters from ringdown waveforms of SMBHs immersed in dark-matter halos.
While adopting the fully relativistic theoretical framework established in~\cite{cardoso2022black}, we generalize the Hernquist profile to a broader family of Dehnen dark-matter halos.
The numerical waveforms are generated using an adapted version of the open-source code from Ref.~\cite{cardoso2022gravitational}.
The data analysis is performed on TDI post-processed data in the context of future space-based GW detection, primarily taking the Taiji mission as a representative example by adopting the mission concept and noise budget from the Taiji Data Challenge~\cite{Du:2025xdq}.
We compute the SNRs of the GW signals obtained via second-order TDI schemes, and Bayesian analysis is employed to quantitatively estimate the underlying parameters of the theoretical models.

The remainder of the paper is organized as follows.
We begin in Sec.~\ref{sec:2} by discussing the theoretical framework, which consists of a central SMBH immersed in a dark matter halo.
We present the coupled master equations governing metric and matter perturbations, along with the specific dark matter profiles, including the Dehnen model and its geometrically corrected version.
The main numerical results are presented in Sec .~\ref {sec:3}.
After specifying the numerical scheme, initial and boundary conditions, we evaluate the SNRs by adopting realistic parameters for the Taiji detector.
These results are then compared using the SNR of the difference waveform for various models of the Dehnen family with respect to the case without a dark matter halo.
Finally, we analyze the performance of a Bayesian analysis carried out for parameter estimation.
The last section is devoted to further discussion and concluding remarks.

\section{The theoretical framework}\label{sec:2}

In this work, we adopt the fully relativistic framework developed in Refs .~\cite {allen1998gravitational, cardoso2022gravitational}.
The gravitational system concerns a spherically symmetric, static black hole embedded in a dark matter halo.
In this framework, the dark matter is modeled as an anisotropic fluid satisfying a barotropic equation of state, with its degrees of freedom minimally coupled to the gravitational sector.
The master equations can be derived by perturbing the metric as well as the matter field.
Existing results indicate that, while the axial gravitational sector is effectively decoupled from the fluid, the polar sector is intrinsically coupled with the matter degrees of freedom. 
As will be elaborated below, for the degrees of freedom associated with the polar sector, the temporal evolution of initial perturbations is governed by a system of three wavelike equations, and the presence of dark matter manifests itself as more resilient fluid modes in the resulting waveforms.
From a physical perspective, the master equation describes the physical process of the late stages of a black hole merger: when the two black holes are sufficiently close, in the close-limit approximation, they can be treated as a single black hole surrounded by a common event horizon. 
Therefore, in this case, the system's evolution can be effectively viewed as a perturbation of a single supermassive black hole. 

\subsection{The master equations of polar gravitational perturbations}

The background metric around a black hole surrounded by a dark matter halo is taken to be the following spherically symmetric form:
\begin{equation}
{\rm d}s^2 = -a(r) {\rm d}t^2 + \frac{{\rm d} r^2}{b(r)} + 
r^2 \left( {\rm d} \theta^2 +\sin^2 \theta {\rm d} \varphi^2 \right) \ ,
\end{equation}
where the metric coefficients $a(r)$ and $b(r)$ are 
functions of $r$ only. Specifically,
\begin{equation}
b(r) = 1 - \frac{2m(r)}{r}  ,
\end{equation}
and  $m(r)$ represents the mass inside radius $r$.

For the metric perturbations, we adopt the Regge-Wheeler gauge~\cite{regge1957stability, zerilli1970effective}, in which the linearized perturbations, viewed as a decomposition in terms of the scalar, vector, and tensor harmonics as the irreducible representations of the SO(3) little group, are required to satisfy a few specific conditions.
In particular, the tensor harmonics are set to zero, and the vector harmonics are taken to be divergence-free~\footnote{Here, we follow Nollert's terminology~\cite{nollert1999quasinormal}, although all of these basis functions are referred to as tensor harmonics by Zerilli~\cite{zerilli1970effective}.
}, namely~\cite{nollert1999quasinormal},
\[
h_{0\phi} = 0 \quad , \quad h_{\phi\phi} = h_{00} \sin^2 \theta ,
\]
\[
\partial_{\phi} \left( \frac{h_{t\phi}}{\sin\theta} \right) + \partial_{\theta} \left( h_{t\theta} \sin\theta \right) = 0 ,
\]
\begin{equation}
\partial_{\phi} \left( \frac{h_{r\phi}}{\sin\theta} \right) + \partial_{\theta} \left( h_{r\theta} \sin\theta \right) = 0 .
\end{equation}
To proceed further, one can decompose the metric perturbations into odd and even parities, also known as the axial and polar gravitational perturbations:
\begin{equation}
\begin{aligned}
h_{\mu\nu}^{(1)\text{axial}}(t,r,\theta,\phi) &= \sum_{\ell=2}^{\infty} \sum_{m=-\ell}^{\ell} \frac{\sqrt{2\ell(\ell+1)}}{r} \left[ i h_1^{\ell m}(t,r) c_{\ell m,\mu\nu}(r,\theta,\phi) - h_0^{\ell m}(t,r) c_{\ell m,\mu\nu}^0(r,\theta,\phi) \right], \\
h_{\mu\nu}^{(1)\text{polar}}(t,r,\theta,\phi) &= \sum_{\ell=2}^{\infty} \sum_{m=-\ell}^{\ell} \left[ a H_0^{\ell m}(t,r) a_{\ell m,\mu\nu}^0(\theta,\phi) - i \sqrt{2} H_1^{\ell m}(t,r) a_{\ell m,\mu\nu}^1(\theta,\phi) \right. \\
&\quad \left. + \frac{H_2^{\ell m}(t,r)}{b} a_{\ell m,\mu\nu}(\theta,\phi) + \sqrt{2} r^2 K^{\ell m}(t,r) g_{\ell m,\mu\nu}(r,\theta,\phi) \right].
\end{aligned}
\end{equation}
with ${h_1^{\ell m}, h_0^{\ell m}, H_0^{\ell m}, H_1^{\ell m}, H_2^{\ell m}, K^{\ell m}}$ are the six remaining free expansion coefficients of the harmonic bases. 

For the GWs, only two out of the six degrees of freedom remain physical after fixing the gauge.
In a vacuum, they satisfy the Regge-Wheeler and the Zerilli equations~\cite{regge1957stability,zerilli1970effective}, respectively. 
Owing to the presence of dark matter, these master equations for the gravity sector suffer further modifications.
In practice, the background dark matter profile is assumed, and Cardoso {\it et al.} specifically employ a Hernquist-type distribution featuring the spike~\cite{cardoso2022black, hernquist1990analytical}:
\begin{equation}
\label{Hernquist}
\begin{aligned}
\rho(r) =
\begin{cases}
\frac{M_{\text{halo}} (a_0 + 2M_{\text{BH}}) \left(1 - \frac{2M_{\text{BH}}}{r} \right)}
{r (r + a_0)^3 (2\pi)},  & r \geq r_{\text{cut}} \\[10pt]
\left[1 - \left(\frac{r_{\text{cut}} - r}{r_{\text{cut}} - r_{\text{zero}} } \right)^4 \right] 
\frac{M_{\text{halo}} (a_0 + 2M_{\text{BH}}) \left(1 - \frac{2M_{\text{BH}}}{r} \right)}
{r (r + a_0)^3 (2\pi)},  & r_{\text{zero}}  \leq r < r_{\text{cut}} \\[10pt]
0,  & r < r_{\text{zero}} 
\end{cases}
\end{aligned}
\end{equation}
and 
\begin{equation}
m(r) = M_{\text{BH}} + \frac{M_{\text{halo}} r^{2}}{(a_{0} + r)^{2}} \left(1 - \frac{2M_{\text{BH}}}{r}\right)^{2} ,
\end{equation}
where $\rho(r)$ is the density profile and $m(r)$ is the mass profile enclosed within the Schwarzschild radius $r$. $M_{\text{BH}}$ represents the mass of the central supermassive black hole, while $M_{\text{halo}}$ and $a_{0}$ denote the total mass and the scale radius of the background Hernquist distribution, respectively. 
This configuration can be found in the open-source numerical code~\cite{GRIT_Code, zenginouglu2011null}.
Besides, the background stress-energy tensor is set to have an anisotropic form~\cite{cardoso2022gravitational} :
\begin{equation}
(T^{(0)\text{env}})^\mu{}_\nu = \text{diag}(-\rho, 0, P_{t} , P_{t} ) \, ,\label{TmunuDMatter}
\end{equation}
where one has assumed a vanishing radial pressure and $P_{t}$ is the tangential pressure.
For different choices of $\rho \left ( r\right )$, the background metric is governed by the field equation, which gives:
\begin{equation}
a' = \frac{2a}{r^2} \frac{m}{1 - 2m/r}, 
\end{equation}
\begin{equation}
P_{t} = \frac{m}{2(r - 2m)} \rho,
\end{equation}
where $a'$ represents the derivative of $a$ with respect to $r$. 
One therefore has
\begin{gather}
a(r) = \left(1 - \frac{2M_{\text{BH}}}{r}\right) e^{\Upsilon},
\\
\Upsilon = -\pi \sqrt{\frac{M_{\text{halo}}}{\xi}} + 2 \sqrt{\frac{M_{\text{halo}}}{\xi}} \arctan \left( \frac{r + a_{0} - M_{\text{halo}}}{\sqrt{M_{\text{halo}} \xi}} \right),
\\
\xi = 2a_{0} - M_{\text{halo}} + 4M_{\text{BH}}.
\end{gather}

Similar to the metric perturbations, one can also expand the linearized deformations of the pressure and density of the dark matter field as follows:
\begin{align}
P_r^{(1)}(t, r, \theta, \phi) &= \sum_{\ell=2}^{\infty} \sum_{m=-\ell}^{\ell} \delta p_{r, \ell m}(t, r) Y_{\ell m} (\theta, \phi) ,\\[10pt]
P_t^{(1)}(t, r, \theta, \phi) &= \sum_{\ell=2}^{\infty} \sum_{m=-\ell}^{\ell} \delta p_{t, \ell m}(t, r) Y_{\ell m} (\theta, \phi) , \\[10pt]
\rho^{(1)}(t, r, \theta, \phi) &= \sum_{\ell=2}^{\infty} \sum_{m=-\ell}^{\ell} \delta \rho_{\ell m}(t, r) Y_{\ell m} (\theta, \phi) .
\end{align}
For a barotropic equation of state, the dark matter's speed of sound is
\begin{equation}
\delta p_{t,r}^{lm}(t,r) = c_{s_{t,r}}^2(r) \delta \rho^{lm}(t,r).
\end{equation}
where $c_{s_r}$ and $c_{s_t}$ are the radial and transverse sound speeds.
We choose the values $c_{s_{t,r}} = (0.9, 0)$ throughout the paper~\cite{zhang2024detecting}.

Owing to its coupling to the matter field, the polar gravitational sector naturally carries the dynamical information about the dark matter content. 
Therefore, the present study will primarily focus on polar perturbations. 
In what follows, we give a brief account of the master equations adopting the notations of~\cite{cardoso2022gravitational, allen1998gravitational}.
For the gravity sector, these degrees of freedom are essentially governed by the linearized Einstein field equations,
\begin{equation}
\delta G_{\mu\nu} = 8\pi \delta T_{\mu\nu} .
\end{equation}
We will utilize the tortoise coordinate $r_*$ to replace the original radial coordinate $r$ in the spatial derivatives
\begin{equation}
\frac{{\rm d}{r}_{*}}{{\rm d}r}=\sqrt{a\left ( r\right )b\left ( r\right )}.
\end{equation}
In the Regge-Wheeler gauge, the degrees of freedom are $H_0$, $H_1$, $H_2$, and $K$, which are associated with the coefficients of three scalar harmonics and one tensor harmonic.
Following~\cite{PhysRevD.42.1884,cardoso2022gravitational, allen1998gravitational}, one introduce $S$ and $\widetilde{H}_1$ to replace $H_0$ and $H_1$
\begin{gather}
H_0(t,r) = K(t,r) + \frac{r}{a} S(t,r), \\
H_1(t,r) = \frac{r}{a} \widetilde{H}_1(t,r).
\end{gather}
Here, for brevity, we omit the $\left ( l,m\right )$ indices for the harmonic basis, and the same applies hereafter.
The resulting equations of motion for $\widetilde{H}_1$, $H_2$, $K$, and $S$ are
\begin{equation}
\frac{\partial \tilde{H}_{1}}{\partial t} = \sqrt{\frac{a}{1-2 m / r}} \frac{\partial S}{\partial r_{*}} + \frac{a}{r} S + \frac{2 a^{2} m}{r^{3}(1-2 m / r)} K,
\end{equation}
\begin{equation}
H_2(t,r) = K(t,r) + \frac{r}{a} S(t,r) ,
\end{equation}
\begin{equation}
\label{HC}
\begin{aligned}
\frac{\partial^2 K}{\partial r_*^2} &= \sqrt{\frac{1-2m/r}{a}} \frac{\partial S}{\partial r_*} + \sqrt{\frac{a}{1-2m/r}} \left( \frac{5m}{r^2} - \frac{2}{r} \right) \frac{\partial K}{\partial r_*} \\
&\quad + a \left[ \frac{\ell(\ell+1)}{r^2} - 8\pi \rho \right] K + \left[ \frac{\ell(\ell+1) + 4}{2r} - \frac{4m}{r^2} - 8\pi r \rho \right] S - 8\pi a \, \delta \rho,
\end{aligned}
\end{equation}
\begin{equation}
\begin{aligned}
-\frac{\partial^2 K}{\partial t^2} + \frac{\partial^2 K}{\partial r_*^2} &+ \frac{2}{r} \sqrt{a(1-2m/r)} \frac{\partial K}{\partial r_*} + \frac{a}{r^2} \left[ 8\pi r^2 \rho + \frac{4m}{r} - \ell(\ell+1) \right] K \\
&= -8\pi a \left( 1 - c_{s_r}^2 \right) \delta \rho + \frac{2}{r} \left( 1 - \frac{2m}{r} - 4\pi r^2 \rho \right) S ,
\end{aligned}
\end{equation}
and
\begin{equation}
\begin{aligned}
-\frac{\partial^2 S}{\partial t^2} + \frac{\partial^2 S}{\partial r_*^2} &+ \frac{a}{r^2} \left[ 4\pi r^2 \rho + \frac{2m}{r} - \ell(\ell+1) \right] S \\
&= \frac{4a^2}{r^4(r-2m)} \left[ 3m(r + 2\pi r^3 \rho) - 7m^2 - 4\pi r^4 \rho \right] K - 16\pi \frac{a^2}{r} \left( c_{s_r}^2 - c_{s_t}^2 \right) \delta \rho.
\end{aligned}
\end{equation}
where the equations for $K$ and $S$ are {\it wavelike}.
Some of these equations impose algebraic relations that, together with other constraints, must be satisfied not only by the initial data but also enforced throughout the time evolution.
We will discuss these constraints in more detail in the next section.

The third wavelike equation is for the matter field, derived from the conservation of the stress-energy tensor.
One finds
\begin{equation}
\begin{aligned}
& -\frac{\partial^{2}\delta\rho}{\partial t^{2}} 
+ c_{s_{r}}^{2}\frac{\partial^{2}\delta\rho}{\partial r_{*}^{2}} 
+ \sqrt{\frac{a}{1-2m/r}} \left[ \frac{2}{r}\left(2c_{s_{r}}^{2}-c_{s_{t}}^{2}\right) 
+ \left(1-5c_{s_{r}}^{2}+4c_{s_{t}}^{2}\right)\frac{m}{r^{2}} 
+ 2\left(1-\frac{2m}{r}\right)c_{s_{r}}c_{s_{r}}^{\prime} \right] \frac{\partial\delta\rho}{\partial r_{*}} \\
& + \frac{a}{1-2m/r} \Bigg\{ \frac{2c_{s_{r}}^{2}-c_{s_{t}}^{2}\left(\ell^{2}+\ell+2\right)}{r^{2}} 
+ \frac{2m}{r^{3}}\left[c_{s_{t}}^{2}\ell\left(\ell+1\right) 
+ \left(1-3c_{s_{r}}^{2}+4c_{s_{t}}^{2}\right)\frac{m}{r}\right] \\
& \quad + 8\pi\rho\left[1+2c_{s_{t}}^{2} 
+ \left(c_{s_{r}}^{2}-4c_{s_{t}}^{2}-1\right)\frac{m}{r}\right] 
+ \frac{2}{r^{2}}\left(1-\frac{2m}{r}\right)(2r-3m)c_{s_{r}}c_{s_{r}}^{\prime} 
- \frac{4}{r}\left(1-\frac{2m}{r}\right)^{2}c_{s_{t}}c_{s_{t}}^{\prime} \Bigg\} \delta\rho \\
& = -\frac{r}{2}\frac{\partial\rho}{\partial r}\sqrt{\frac{1-2m/r}{a}}\frac{\partial S}{\partial r_{*}} 
- \frac{1}{2r}\sqrt{\frac{a}{1-2m/r}} \left[ (m-r)\frac{\partial\rho}{\partial r} 
- \left(\frac{m}{r}+4\pi r^{2}\rho\right)\frac{\rho}{1-2m/r} \right] \frac{\partial K}{\partial r_{*}} \\
& - \frac{1}{2} \left\{ \left[8\pi r\rho - \frac{m}{r^{2}}\ell\left(\ell+1\right)\right]\frac{\rho}{1-2m/r} 
+ \frac{\partial\rho}{\partial r} \right\} S - \frac{a}{2r^{2}} \left\{ \left[8\pi r^{2}\rho - \ell\left(\ell+1\right)\frac{m}{r}\right]\frac{\rho}{1-2m/r} 
+ 4m\frac{\partial\rho}{\partial r} \right\} K.
\end{aligned}
\end{equation}

\subsection{Dark-matter halo profiles}\label{sec:2b}

In addition to the Hernquist-type dark-matter profile described by Eq.~(\ref{Hernquist}), we also consider two other representative models belonging to the Dehnen family. 
Introduced by Dehnen in 1993~\cite{dehnen1993family}, the latter is commonly used to describe the density distribution of dark matter in galaxy halos in studies of large-scale structure formation and galaxy evolution.
Compared with the Navarro-Frenk-White (NFW) profile~\cite{navarro1997universal,navarro1996structure}, it offers a simple, analytic, flexible form that captures how dark-matter density varies from the center to the outer regions. 
By introducing an adjustable parameter $\gamma$, 
\begin{equation}
\label{Dehnen}
\rho(r)=\frac{(3-\gamma)M_{\text{halo}}}{4\pi}\,\frac{a_0}{r^{\gamma}(r+a_0)^{4-\gamma}} \, ,
\end{equation}
the model encompasses a range of dark-matter configurations.

\begin{figure}[htbp] 
    \centering
    \includegraphics[width=0.8\textwidth]{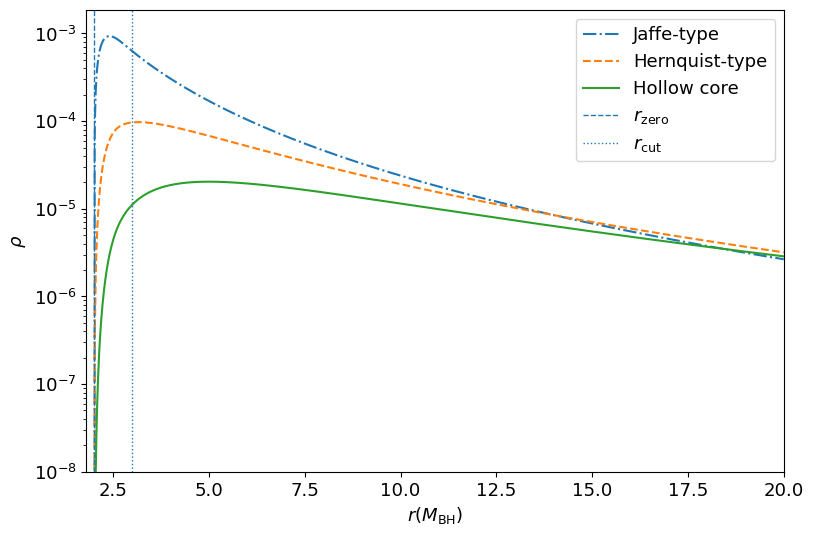}
    \caption{Geometrically corrected density profiles of the Dehnen dark-matter family with a given halo mass $M_\text{halo}$, for different shapes governed by the parameter $\gamma = 0, 1$, and $2$. 
    One assumes $r_{\text{zero}} = 2.5 M_{\text{BH}}$, $r_{\text{cut}}= 3 M_{\text{BH}}$, $M_{\text{halo}}=0.1M_{\text{BH}}$, and $a_0=10M_{\text{halo}}$, and the radial coordinate $r$ is also expressed in the units of $M_{\text{BH}}$.}
    \label{fig:31}
\end{figure}

Specifically, in this work, we consider, in addition to a Hernquist-type profile $(\gamma = 1)$, a Jaffe-type profile $(\gamma = 2)$, and a hollow-core profile $(\gamma = 0)$.
Similar to Hernquist's case elaborated in~\cite{cardoso2022black}, the corresponding mass functions are:
\begin{equation}
m(r)=M_{\text{BH}} + \frac{M_{\text{halo}}\,r}{a_{0} + r}\left(1 - \frac{2M_{\text{BH}}}{r}\right) \ \ \ (\text{for} \ \gamma  = 2) 
\end{equation}
and 
\begin{equation}
m(r)=M_{\text{BH}} + \frac{M_{\text{halo}}\,r^3}{(a_{0} + r)^3}\left(1 - \frac{2M_{\text{BH}}}{r}\right)^3\ \ \ (\text{for} \ \gamma  = 0) .
\end{equation}
Moreover, one needs to ensure that the dark matter halo is well-behaved at the horizon, which is implemented by introducing a truncation at $r=r_\text{zero}$.  
The resulting density profiles read
\begin{equation}
\rho(r) = 
\begin{cases}
\frac{ {M_{\text{halo}}} \cdot (a_0 + 2 {M_{\text{BH}}}) }{ r^{2} (r + a_0)^{2}  4\pi } & \text{if } r \geq r_{\text{cut}} \\
\left[ 1 - \left( \frac{ r_{\text{cut}} - r }{ r_{\text{cut}} - r_{\text{zero}}  } \right)^4 \right]  \frac{ M_{\text{halo}} (a_0 + 2 M_{\text{BH}}) }{ r^{2} (r + a_0)^{2}  4\pi } & \text{if } r_{\text{zero}}  \leq r < r_{\text{cut}} \\
0 & r < r_{\text{zero}} 
\end{cases}\ \ \ (\text{for} \ \gamma  = 2) ,
\end{equation}
and
\begin{equation}
\rho(r) = 
\begin{cases}
\frac{3 M_{\text{halo}} (a_0 + 2 M_{\text{BH}}) \left(1 - \frac{2 M_{\text{BH}}}{r}\right)^2}{4\pi (r + a_0)^4} & \text{if } r \geq r_{\text{cut}} \\
\left[ 1 - \left( \frac{ r_{\text{cut}} - r }{ r_{\text{cut}} - r_{\text{zero}}  } \right)^4 \right] \frac{3 M_{\text{halo}} (a_0 + 2 M_{\text{BH}}) \left(1 - \frac{2 M_{\text{BH}}}{r}\right)^2}{4\pi (r + a_0)^4} & \text{if } r_{\text{zero}}  \leq r < r_{\text{cut}} \\
0 & \text{if } r < r_{\text{zero}} 
\end{cases}\ \ \ (\text{for} \ \gamma  = 0) .
\end{equation}
Here, when $r \geq r_{\text{cut}}$, $\rho(r) = {m}^{\prime}\left ( r\right )$. Compared to the original dark matter profile Eq.~(\ref{Dehnen}), the geometric correction factor $\left(1 - {2M_{\text{BH}}}/{r}\right)$ is introduced for the mass $m(r)$, along with a smooth cutoff imposed on the density $\rho(r)$ within $r_{\text{cut}}$.
These introduced modifications are made with reference to Eq.~(\ref{Hernquist}) and aim to ensure that the Dehnen dark matter family conforms to the geometric properties around a black hole, such as the absence of matter density at the black hole horizon, while preserving the fundamental power-law characteristics of the Dehnen family's matter density with respect to the radius $r$. Provided these requirements are satisfied, we can appropriately choose the exponent of the modification factor for $m(r)$, which is convenient for obtaining the analytical form of the metric~\cite{cardoso2022black}.
The density profiles of the three scenarios considered here are illustrated in Fig.~{\ref{fig:31}}, by assuming $r_{\text{zero}} = 2.5 M_{\text{BH}}$, $r_{\text{cut}}= 3 M_{\text{BH}}$, $M_{\text{halo}}=0.1M_{\text{BH}}$, and $a_0=10M_{\text{halo}}$.

For these scenarios, analytical forms of the metric parameter $a$ are accessible, and found to be:
\begin{equation}
a_{\gamma  = 2}=\exp\!\left[
\frac{2\,M_{\text{halo}}\;\ln \left(\frac{r - 2M_{\text{BH}} }{a_{0} - 2M_{\text{BH}} + r }\right)}
{a_{0} + 2M_{\text{BH}} - 2M_{\text{halo}}}\right]\left(1-\frac{2M_{\text{BH}}}{r}\right)
\end{equation}
and 
\begin{equation}
a_{\gamma  = 0}=\exp\left \{-2M_{\text{halo}}  \operatorname{RootSum}[\zeta (x),\xi (x)]\right \}\left(1-\frac{2M_{\text{BH}}}{r}\right),
\end{equation}
where $\operatorname{RootSum}[\cdots]$ denotes the sum of all roots of the functions given in the brackets, and 
\[
\zeta (x)=2a_0^{2}M_{\text{halo}} + 8a_0M_{\text{BH}}M_{\text{halo}} + 8M_{\text{BH}}^{2}M_{\text{halo}} 
- 4a_0M_{\text{halo}}x - 8M_{\text{BH}}M_{\text{halo}}x + 2M_{\text{halo}} {x}^{2} -{x}^{3} ,
\]
\[
\xi (x)=\frac{ 
    a_0 \ln(a_0 + r - x) + 2 M_{\text{BH}} \ln(a_0 + r - x) - \ln(a_0 + r - x) \, x 
}{ 
    4 a_0 M_{\text{halo}} + 8 M_{\text{BH}} M_{\text{halo}} - 4 M_{\text{halo}} x + 3 x^2 
}.
\]

\section{Numerical results}\label{sec:3}

\subsection{The initial and boundary conditions and resulting waveforms}\label{sec:3a}

In the previous section, we discussed the equations that govern perturbations. 
By performing a linear perturbation analysis of the Einstein field equations and conservation of the stress-energy tensor, we obtain a set of three coupled wavelike equations, which are hyperbolic partial differential equations governing the evolution of the metric perturbations and dark matter fluid, together with a few constraints. 
Given the complexity of the system, analytical solutions are generally unattainable, and numerical techniques are required.
Before proceeding with numerical integration, it is crucial to rigorously define appropriate initial data and boundary conditions, as these are fundamental to the well-posedness of the Cauchy problem~\cite{Kind_1993, PhysRevD.46.4289}.

To approximate the physical scenario, we impose a Dirichlet condition at the inner boundary near the black hole horizon for the metric perturbations and place the inner boundary sufficiently far away from the extraction radius such that they cannot affect the evaluation of relevant quantities.
As for the outer boundary, we assume that outgoing waves can propagate freely to spatial infinity, and thus no explicit boundary condition is imposed there. 
For the perturbations in the matter field, we will prescribe a Dirichlet boundary condition at $r_{\text{zero}} $. 

As for the initial data, defining acceptable initial data for the evolution problem is not a trivial task. 
The construction of initial data fundamentally hinges on the Hamiltonian constraint, as expressed in Eq.~(\ref{HC})~\cite{arnowitt2008dynamics,cardoso2022gravitational}. 
A single constraint is insufficient to specify the complete initial configuration of the three coupled fields. 
A physically rigorous construction of initial data would require a full dynamical model of how the black hole merger interacts with the environment, especially the generation and evolution of matter-field perturbations. 
The resulting initial perturbations would then need to be consistent with the dynamic equations employed in this work while serving as initial data for the evolution equations. 
However, as the central objective here is to explore how the dark matter distribution influences the GW waveforms and whether the resultant GWs may enable the inference of environmental properties, such modeling lies beyond the scope of this paper.
Instead, we will adopt a time-symmetric initial data, 
\begin{eqnarray}
\partial_t {K}|{}_{ t=0}=\partial_t {S}|{}_{ t=0}=\partial_t {\delta \rho }|{}_{ t=0}=0 .
\end{eqnarray} 
Such a choice ensures that the time derivative of the Hamiltonian constraint is automatically satisfied. 

\begin{figure}[htbp] 
    \centering
    \includegraphics[width=0.8\textwidth]{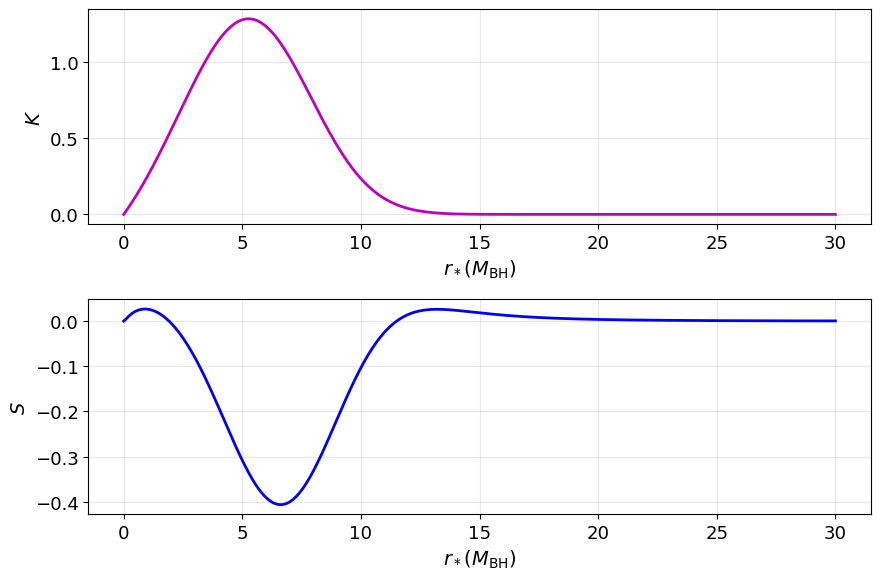}
    \caption{An example of the first class of initial conditions, in which the initial matter density perturbation is set to zero. 
    The initial perturbations of the fields $K$ and $S$ are shown in the tortoise coordinate, where the amplitude is chosen as $A=1$ in Eq.~\eqref{KiniC} for demonstration purposes.} 
    \label{fig:12}
\end{figure}

\begin{figure}[htbp] 
    \centering
    \includegraphics[width=0.8\linewidth]{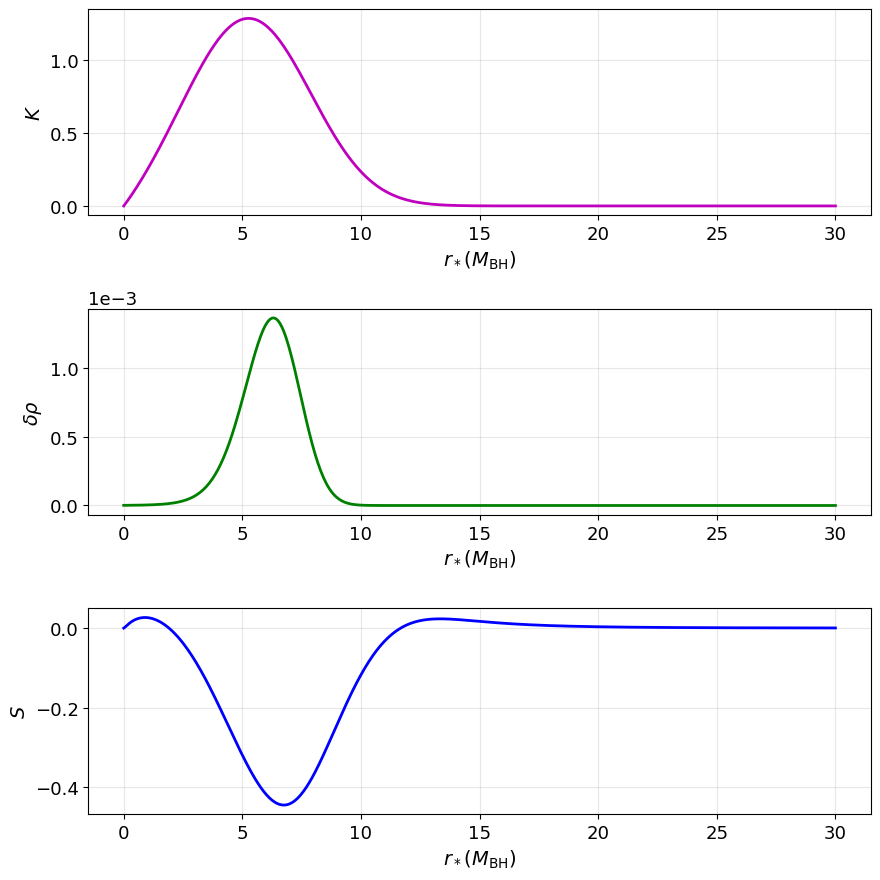}
    \caption{An illustration of the second class of initial conditions.
    It is assumed that the destruction of the cusp leads to a large initial matter-density perturbation. 
    The initial perturbations of the fields $K$, $S$, and $\delta\rho$ are presented in the tortoise coordinate.}
    \label{fig:9}
\end{figure}

Following prior work~\cite{allen1998gravitational}, to satisfy the Hamiltonian constraint, we explore two distinct classes of initial data sets consisting of a set of simplified conditions for $K$, $S$, and $\rho$, reflecting different physical scenarios.
The first scenario corresponds to a configuration with no initial perturbation in the matter density. 
This scenario can be interpreted as the scattering of GWs by the effective potential of a black hole embedded in a dark matter distribution.
The second scenario is more involved. 
When the merger process is sufficiently violent, the impact on the dark-matter halo is sizable.
In that case, it is no longer justified to treat the dark matter as a static background.  
In particular, if the black holes recoil after the merger and acquire a net kick velocity, the spike of the dark-matter distribution might be destroyed along the direction of motion~\cite{merritt2002dark, Merritt2004}. 
We therefore explicitly consider the impact on the dark matter halo as a matter field perturbation that constitutes the initial conditions, which propagates together with the GWs produced by the merger. 
Given that the initial density perturbation may be highly non-trivial, we therefore adopt the following form of simplified initial data that is physically plausible: the initial density perturbation $\delta\rho$ is modeled as a wave packet whose overall shape is determined by the dark-matter spike profile. At this point, the main component of the dark matter halo will act as the initial matter perturbation, coupling with the GWs and co-evolving, rather than serving as a static background for the evolution.
Furthermore, the initial spatial profile of $K$ is prescribed by the Gaussian-like form
\begin{equation}
K\left ( {r}^{*}\right )=A{r}^{*}\exp \left[ -({r}^{*}-{r}_{0})^{2}/{\sigma  }^{2}\right ],\label{KiniC}
\end{equation}
where $A$ and $\sigma$ characterize the amplitude and the width of the wave packet, respectively.
As will be elaborated below, the initial amplitude $A$ could be determined by a specific waveform-generation package that incorporates the properties of the underlying gravitational system, such as those associated with a merger process.
For the width parameter, we adopt $\sigma \approx 2M_{\text{BH}}$, following Ref.~\cite{andersson1995excitation}.
Subsequently, the Hamiltonian constraint, together with the prescribed forms of $K$ and $\delta\rho$, determines the initial spatial profile of $S$.
The resulting two types of initial data are illustrated in Figs.~\ref{fig:12} and~\ref{fig:9}.

\begin{figure}[htbp] 
    \centering
    \includegraphics[width=0.8\textwidth]{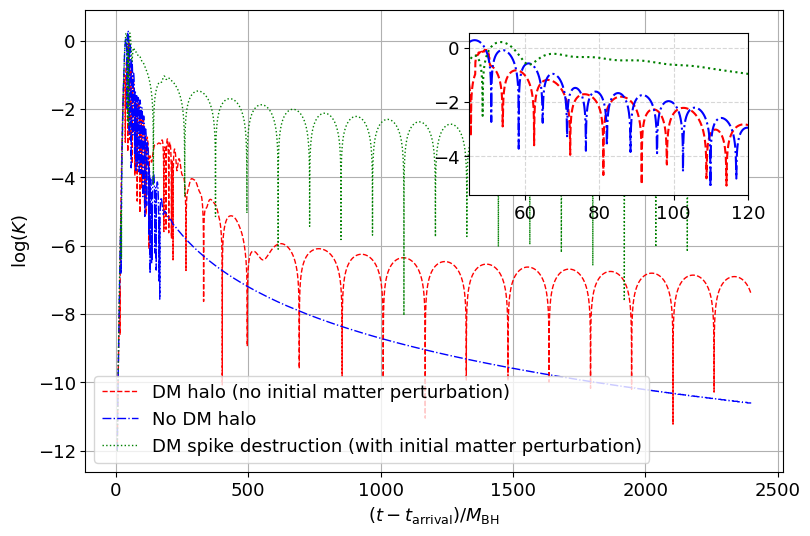}
    \caption{Temporal evolutions of the metric perturbation $K$ for three distinct scenarios. 
    No DM halo: this label corresponds to the blue dash-dotted curve and indicates the reference case in which the black hole evolves in vacuum, without any surrounding dark-matter halo; DM halo (no initial matter perturbation): this label corresponds to the red dashed curve and represents the case where a dark-matter halo is present, but no initial perturbation is imposed on the matter field; DM spike destruction: this label corresponds to the green dotted curve and denotes the case where a dark-matter halo is initially present but subsequently destroyed, thereby supplying the initial matter-density perturbations.
    For the dark matter halo, we adopt the Jaffe model with the parameters $M_{\text{halo}}=1M_{\text{BH}},a_0=100M_{\text{halo}}$.
    The calculations employ the initial conditions given in Figs.~\ref{fig:12} and~\ref{fig:9}.}
    \label{fig:40}
\end{figure}

Given the initial conditions described above, one numerically evolves the coupled system of three hyperbolic equations using the Lax-Wendroff method in the time domain. 
The extraction radius is set at $1000{M}_{\text{BH}}$.
In the absence of a dark matter halo (i.e., setting the halo mass to zero), the simulation reproduces the familiar scenario of metric perturbations scattering off a black hole effective potential, whose magnitude typically falls exponentially in time. 
As shown by the blue dash-dotted curve in Fig.~\ref{fig:40}, the gravitational waveform exhibits the typical ring-down phase followed by an inverse power-law tail, leading to rapid decay of the perturbation. 
When a dark-matter halo is present, the GWs are simultaneously scattered by both the dark-matter potential and the black-hole potential.  While the ring-down phase persists, the power-law tail is replaced by an oscillatory slow decay, as presented by the green dotted and red dashed curves.
This occurs because the GW itself excites a density wave in the matter; the two fields become coupled and propagate together, continuously exchanging energy. 
If the central dark-matter spike has been perturbed, the process is also seeded by a initial pulse in the matter field.
As a result, the coupled evolution of the matter density wave and the GW is observed at an earlier stage. 
Moreover, when matter is initially perturbed, the energy dissipation is predominantly mediated by fluid-type oscillations known as the $f$-mode~\cite{allen1998gravitational, andersson1998towards}.

\begin{figure}[htbp] 
    \centering
    \includegraphics[width=0.8\textwidth]{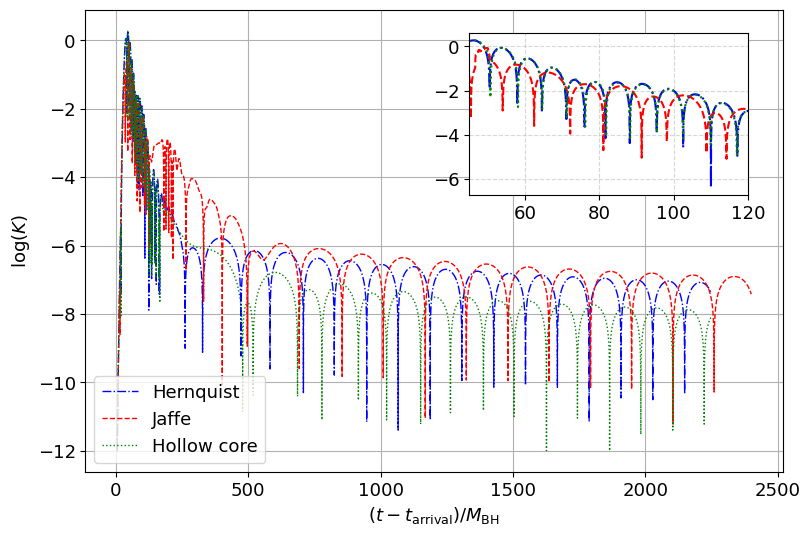}
    \caption{Temporal evolutions of the metric perturbation $K$ for the three Dehnen-family dark-matter models. Hernquist: this label corresponds to the blue dash-dotted curve; Jaffe: this label corresponds to the red dashed curve; hollow core: this label corresponds to the green dotted curve.
    Three scenarios adopt the same initial conditions for $K$ and adhere to the Hamiltonian constraints under their respective density profiles,
    each using dark-matter parameters $M_{\text{halo}}=1M_{\text{BH}},a_0=100M_{\text{halo}}$.}
    \label{fig:41}
\end{figure}

Furthermore, we observe that the damping characteristics and decay rates differ significantly among the various dark-matter profiles.  
In Fig.~\ref{fig:41}, we illustrate the scattering of GWs by the three Dehnen-family dark-matter environments.
It is evident that the sharper the central spike, the earlier fluid modes dominate the evolution, and the larger their amplitudes.  
Nonetheless, the GW signals carried by these fluid modes are much weaker than those of the ring-down stage.  
Therefore, the actual detectable signal strength, especially the response in a large-scale interferometer, requires a dedicated follow-up study.

\subsection{Response in space-based GW interferometers}\label{sec:3b}

Statistical inference analyses often rely on accurate modeling of both the signal and the noise.
As mentioned in Sec .~\ref {sec:intro}, for space-based detections, the extraction of the properties of the GW source is performed using the data that has been processed using the TDI algorithm.
Specifically, in this section, we first model the response of a single laser link to an incident GW signal and instrumental noise, and then construct the corresponding TDI combinations.
As a starting point, the GW data obtained from numerical calculations are conveniently expressed in the Regge--Wheeler gauge.
However, from the perspective of a detector, the measured GW signals correspond to asymptotically plane waves, which are naturally described in the transverse-traceless (TT) gauge. 
These two descriptions are related through a transformation, whose explicit form reads~\cite{nagar2005gauge,PhysRevD.71.104003}
\begin{gather}
\left(h_{+}-\text{i} h_{\times}\right)_{\ell m}^{(\text{polar})} =\frac{1}{r}\sqrt{\frac{(\ell+2)!}{(\ell-2)!}} \Psi_{\ell m}^{(\text{polar})} {}_{-2}Y_{\ell m}(\theta, \phi)+\mathcal{O}\left(\frac{1}{r^{2}}\right)
\end{gather}
where $\Psi_{\ell m}^{(\text{polar})}$ is the Zerilli function:
\begin{equation}
\Psi_{\ell m}^{(\text{polar})} \equiv \frac{r \left(K+{r S}/{a} \right)}{\Lambda [r(\Lambda-2)+6 M_{\text{BH}}]} ,
\end{equation}
where $\Lambda\equiv \ell(\ell+1)$. 
For simplicity, in this work, we retain only the dominant (2, 2) and (2, 1) multipoles. ${}_{-2}Y^{\ell m}(\theta,\phi)$ is the spin-weighted spherical harmonics.

Space-borne detectors such as Lisa and Taiji consist of three spacecraft (S/Cs), and laser beams are exchanged between each pair of them. 
The latter is referred to as laser links, also known as arms, denoted by a pair of indices, $ij \in \mathcal{I}_2 \equiv \{12, 23, 31, 21, 32, 13\}$, corresponding to the two S/Cs that emit and receive the laser beam.
One first analyzes the measured GW signals with respect to a single arm by projecting the signals onto a given laser link.
In practice, one measures the Doppler shift of these laser beams, which potentially records any frequency modulations induced by GWs.
Such ``single-arm'' Doppler observable, denoted as $\eta_{ij}(t)$, can be written as the superposition of a signal term $y_{ij}(t)$ and a noise term $n_{ij}(t)$~\cite{Otto_thesis,Hartwig_thesis,LISA_instrument,Pchannel} 
\begin{equation}\label{eq:single_arm_data}
    \eta_{ij}(t) = y_{ij}(t) + n_{ij}(t),  
\end{equation}
where the relative frequency shift $y_{ij}(t)$ induced by GW reads
\begin{eqnarray}
y_{ij}(t) &\equiv& \frac{\nu_{\rm receive} - \nu_{\rm send}}{\nu_{\rm send}} \nonumber \\
&\approx& \frac{1}{2\left(1-\bm{\hat{k}} \cdot \bm{\hat{n}}_{ij}(t)\right)}\left[H_{ij}\left(t - d_{ij}(t) - \frac{\bm{\hat{k}} \cdot \bm{R}_j(t)}{c}\right) \right. \nonumber 
 \left.- H_{ij}\left(t - \frac{\bm{\hat{k}} \cdot \bm{R}_i(t)}{c}\right)\right] ,
\end{eqnarray}
where $\bm{R}_i(t)$ represents the spatial position of S/C$_i$ in the solar system barycenter (SSB) frame, $d_{ij}(t)$ is the light travel time from S/C$_j$ to S/C$_i$, and $\bm{\hat{k}}$ is the unit vector representing the propagation direction of the GW, and $\bm{\hat{n}}_{ij}(t)$ is the unit vector along this arm. 
The functions $H_{ij}(t)$ denotes the projection of GW tensor on arm $ij$, which takes the following form~\cite{Creighton:2011zz} 
\begin{equation}
    H_{ij}(t) \equiv h_+(t) \zeta_{+, ij}(t) + h_\times(t) \zeta_{\times, ij}(t),
\end{equation}
where the antenna functions $ \zeta_{+, ij}$ and $\zeta_{\times, ij}$ depend on the ecliptic longitude $\lambda$ and ecliptic latitude $\beta$ of the GW source, as well as the polarization angle $\psi$: 
\begin{eqnarray}
\zeta_{+, ij}(t) &=& \cos (2\psi) \xi_{+, ij}(t) + \sin (2\psi) \xi_{\times, ij}(t), \\ 
\zeta_{\times, ij}(t) &=& -\sin (2\psi) \xi_{+, ij}(t) + \cos (2\psi) \xi_{\times, ij}(t), \\
\xi_{+, ij}(t) &=& \left[\bm{\hat{n}}_{ij}(t) \cdot \bm{\hat{u}}\right]^2 - \left[\bm{\hat{n}}_{ij}(t) \cdot \bm{\hat{v}}\right]^2, \\ 
\xi_{\times, ij}(t) &=& 2 \left[\bm{\hat{n}}_{ij}(t) \cdot \bm{\hat{u}}\right]  \left[\bm{\hat{n}}_{ij}(t) \cdot \bm{\hat{v}}\right],
\end{eqnarray}
where unit vectors $\bm{\hat{u}}$, $\bm{\hat{k}}$, and $\bm{\hat{v}}=\bm{\hat{u}}\times \bm{\hat{k}}$ are 
\begin{eqnarray}
    \bm{\hat{u}} &=& \left[\sin \lambda, - \cos \lambda, 0 \right], \\ 
    \bm{\hat{v}} &=& \left[- \sin \beta \cos \lambda, - \sin \beta \sin \lambda, \cos \beta \right], \\ 
    \bm{\hat{k}} &=&  -\left[ \cos \beta \cos \lambda, \cos \beta \sin \lambda, \sin \beta \right].
\end{eqnarray}
In what follows, our analysis will be primarily carried out by adopting the setup and parameters associated with the Taiji detector, specifically those involved in the Taiji Data Challenge~\cite{Du:2025xdq}. 

Regarding the noise term $n_{ij}(t)$, in theory, the laser noise will be suppressed by the TDI algorithm, as elaborated below, to a negligible level.
As a result, the dominant residual noise sources are the optical metrology system (OMS) noises $N_{ij}$ and test-mass acceleration (ACC) noises $\delta_{ij}$.
According to the baseline design of Taiji, the ``nominal'' power spectral densities (PSDs) of these noises take the forms 
\begin{eqnarray}\label{eq:noise_component_PSDs}
    S_{{\rm OMS}}(f) &=& A_{{\rm OMS}}^2 \left(\frac{2\pi f}{c}\right)^2 \left[1 + \left(\frac{2 \ {\rm mHz}}{f}\right)^4\right], \nonumber \\
    S_{{\rm ACC}}(f) &=& A_{{\rm ACC}}^2 \left(\frac{1}{2\pi f c}\right)^2  \left[1 + \left(\frac{0.4 \ {\rm mHz}}{f}\right)^2\right]  \left[1 + \left(\frac{f}{8 \ {\rm mHz}}\right)^4\right], 
\end{eqnarray}
where $A_{{\rm OMS}} = 8 \times 10^{-12} \ \text{m}/\sqrt{\text{Hz}}$ and $A_{{\rm ACC}} = 3 \times 10^{-15} \ \text{m}/\text{s}^2/\sqrt{\text{Hz}}$.
The OMS and ACC noises dominate at the high and low frequencies, respectively. 
For simplicity, we assume that the OMS and ACC noises possess identical PSDs for all $ij$ tuples and are statistically independent of one another.

In general, the TDI observables can be cast into a unified form:
\begin{equation}\label{eq:general_TDI}
    {\rm TDI} \ = \sum_{ij \in \mathcal{I}_2} \textbf{P}_{ij} \eta_{ij},
\end{equation}
where $\textbf{P}_{ij}$ is the polynomial of the delay operators.
As an example, the second-generation Michelson combination $X_2$ is constituted by the following coefficients $\textbf{P}_{ij}$,
\begin{eqnarray}\label{eq:X2_combination}
    \textbf{P}_{12} &=& 1 - \textbf{D}_{131} - \textbf{D}_{13121} + \textbf{D}_{1213131}, \nonumber \\ 
    \textbf{P}_{23} &=& 0, \nonumber \\ 
    \textbf{P}_{31} &=& -\textbf{D}_{13} + \textbf{D}_{1213} + \textbf{D}_{121313} - \textbf{D}_{13121213}, \nonumber \\ 
    \textbf{P}_{21} &=& \textbf{D}_{12} - \textbf{D}_{1312} - \textbf{D}_{131212} + \textbf{D}_{12131312}, \nonumber \\ 
    \textbf{P}_{32} &=& 0, \nonumber \\ 
    \textbf{P}_{13} &=& -1 + \textbf{D}_{121} + \textbf{D}_{12131} - \textbf{D}_{1312121}.  
\end{eqnarray}
In the above expressions, the delay operator $\textbf{D}_{ij}$ is defined as $\textbf{D}_{ij} f (t) \equiv f\left(t - d_{ij}(t) \right)$, when acted on an arbitrary function of time  $f(t)$, and $\textbf{D}_{i_1i_2i_3 ...} f(t) \equiv \textbf{D}_{i_1i_2}\textbf{D}_{i_2i_3}...f(t)$. 
The combinations of $Y_2$ and $Z_2$ channels can be readily obtained from Eqs.~\eqref{eq:X2_combination} by applying the permutation $1 \rightarrow 2 \rightarrow 3 \rightarrow 1$. 
Furthermore, in this work, we consider a simplified scenario where the detector arms are of equal length, a convention extensively adopted in prior studies (e.g., Refs.~\cite{Littenberg:2023xpl,Katz:2024oqg,Deng:2025wgk,Strub:2024kbe} based on LISA Data Challenge~\cite{Baghi:2022ucj}).
Our specifications for the models of detector orbit and noise enable the definition of three noise-orthogonal TDI channels~\cite{Prince:2002hp}, including 
the ``signal'' channels $A_2, E_2$ and  the ``null'' channel $T_2$, defined as 
\begin{eqnarray}\label{eq:AET_from_XYZ}
    A_2 &=& \frac{Z_2 - X_2}{\sqrt{2}}, \\
    E_2 &=& \frac{X_2 - 2Y_2 + Z_2}{\sqrt{6}}, \\
    T_2 &=& \frac{X_2 + Y_2 + Z_2}{\sqrt{3}}. 
\end{eqnarray}

We note that the TDI scheme is expected to suppress the laser frequency noise to below the inevitable secondary noise floor.
Meanwhile, the resulting signals, in terms of TDI combinations of $\eta_{ij}$, become more complicated, as they are now superpositions of time-shifted data streams from different arms.
Therefore, specific TDI combinations should be accounted for when simulating the measured data streams to achieve a more realistic treatment.
Below, we take a specific merger event as an example. We simulate the characteristic gravitational-wave strain for these three combinations using the generated gravitational waveforms, and perform parameter extraction with data from the three combination channels.

For the present study, we consider a possible scenario in which the GW emission originates from a merger event.
To this end, we adopt the parameters listed in Table~\ref{tab:params} and employ the BBHx package to generate the GW waveform~\cite{PhysRevD.102.023033,PhysRevD.105.044055}.
Specifically, BBHx incorporates the IMRPhenom family of waveform models, and the resulting frequency-domain signals are converted into time-domain responses through the Inverse Fast Fourier Transform (IFFT).
From the resulting time-domain waveform, we identify the evolution from the adiabatic inspiral stage to the onset of the ringdown phase.
In the present analysis, the merger-ringdown transition is defined by the instant at which the strain amplitude reaches its maximum, signaling the completion of the merger process.
By isolating the waveform segment preceding the ringdown phase, we extract the peak strain together with the associated orbital dynamics prior to the formation of the final black hole.
The corresponding peak strain is then assigned to the amplitude $A$ introduced previously in Eq.~\eqref{KiniC}.

\begin{table}[htbp]
    \centering
    \caption{Binary system parameters}
    \label{tab:params}
    \begin{tabular}{l c c}
        \hline
        Parameter & Value & Unit \\
        \hline
        Chirp mass & $2.0\times 10^{4}$ & $M_{\odot}$ \\
        Mass ratio & 0.0150 & -- \\
        Coalescence time & 29.9999 & day \\
        Coalescence phase & 3.2535 & rad \\
        Luminosity distance & 5520.52 & Mpc \\
        Inclination & 1.4495 & rad \\
        Ecliptic longitude & 3.0328 & rad \\
        Ecliptic latitude & -1.1818 & rad \\
        Polarization & 0.2409 & rad \\
        SNR of inspiral and merger phase & 356 & -- \\
        Dark matter halo mass & 1.0 & $M_{\text{BH}}$ \\
        Characteristic length of halo  & 100.0 & $M_{\text{BH}}$ \\
        Tangential sound speed & 0.9 & -- \\
        \hline
    \end{tabular}
\end{table}

To assess whether a GW signal can be reliably extracted after processing the TDI algorithm, two quantities, the SNR and the characteristic strain, play a central role. 
They are thus analyzed in the following.
The SNR is estimated by an integral over the sensitive frequency band of the detectors
\begin{equation}\label{avSNR}
\text{SNR}^{2} =\left\langle {h}_{\text{TDI}} \mid {h}_{\text{TDI}}\right\rangle
= 4\int_{f_{\rm min}}^{f_{\rm max}} \frac{|\tilde{h}_{\text{TDI}}(f)|^{2}}{S_{n, \text{TDI}}(f)}\, {\rm d}f ,
\end{equation}
where for the Taiji detector, one takes $f_{\rm min}= 5\times 10^{-4} \text{Hz}$ and $f_{\rm max}=5\times 10^{-1} \text{Hz}$~\cite{ruan2020taiji}.

\begin{figure}[htbp] 
    \centering
    \includegraphics[width=0.8\textwidth]{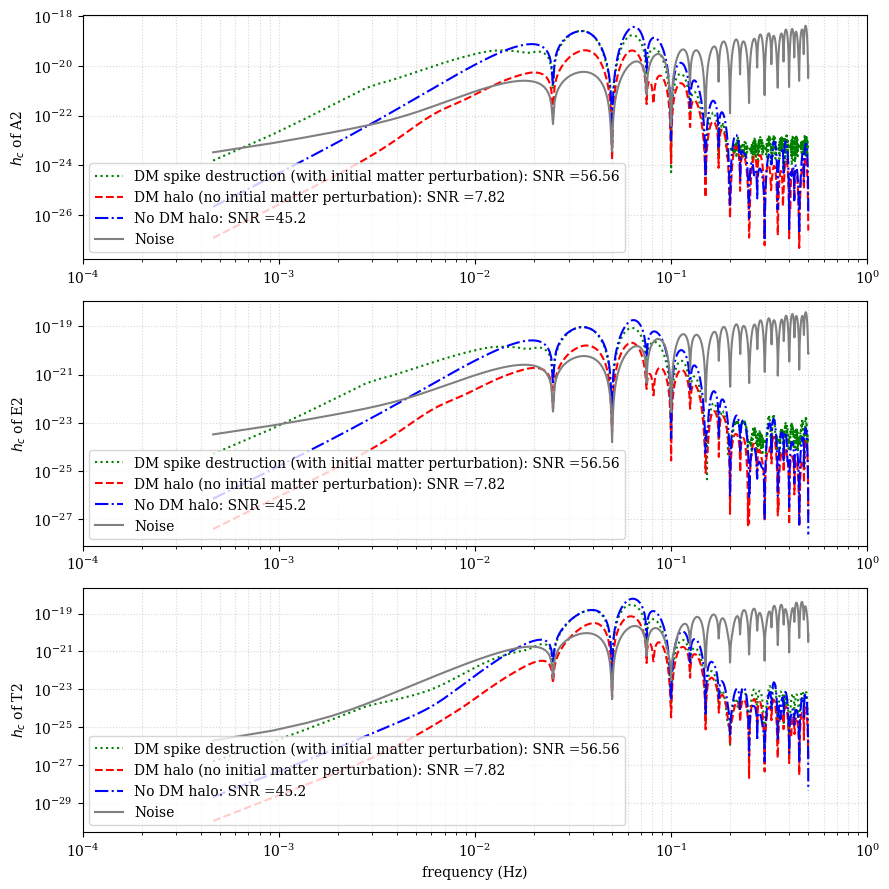}
    \caption{Characteristic strains of GWs in the $A_2$, $E_2$ and $T_2$ channels for three distinct scenarios. 
    The gray solid line represents the noise in the three signal channels after TDI processing. The remaining curves represent the vacuum case (blue dash-dotted), the dark matter halo without initial perturbation (red dashed), and the dark matter spike destruction (green dotted), representing their respective response results in GW detection. The amplitudes are normalized based on the merger event parameters provided in Table~\ref{tab:params}.}
    \label{fig:42}
\end{figure}

\begin{figure}[htbp] 
    \centering
    \includegraphics[width=0.8\textwidth]{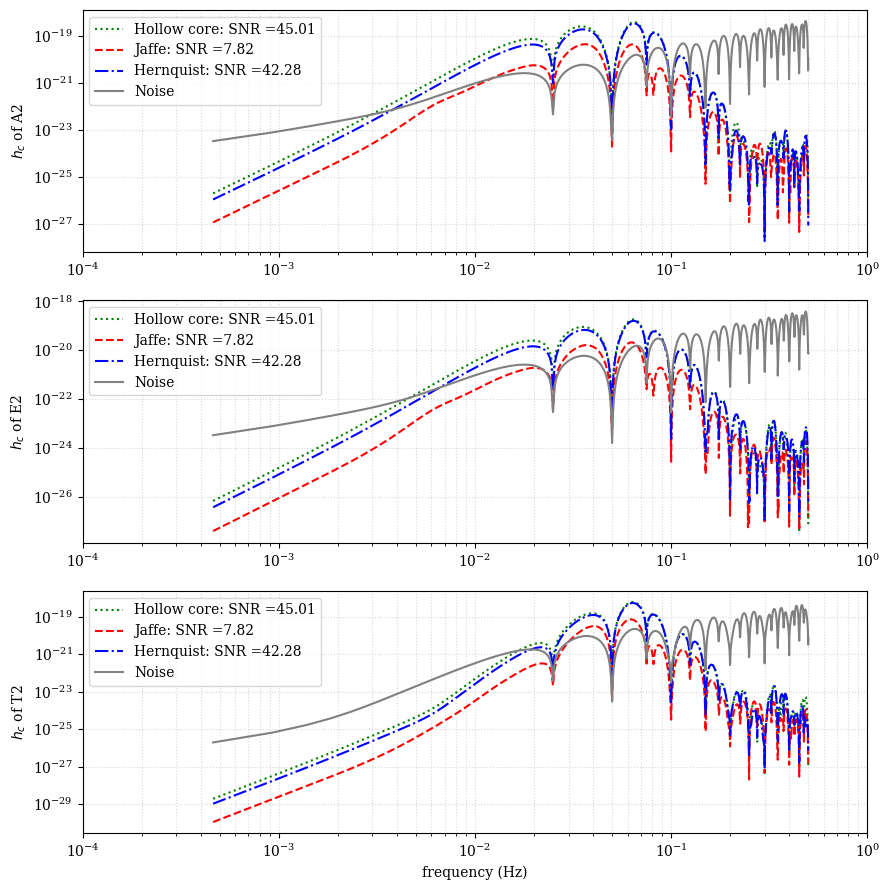}
    \caption{Characteristic strains of GWs in the $A_2$, $E_2$ and $T_2$ channels for the three Dehnen-family dark-matter models.  
    The gray solid line represents the noise in the three signal channels after TDI processing.
    The remaining curves represent the Hernquist scenario (blue dash-dotted), Jaffe scenario (red dashed), and hollow core scenario (green dotted), representing their respective response results in GW detection.}
    \label{fig:43}
\end{figure}

In Figs.~\ref{fig:42} and~\ref{fig:43}, we present the characteristic strains and SNRs that correspond to the initial conditions of the three different scenarios discussed in the previous section. 
From the characteristic strains shown in Fig.~{\ref{fig:42}}, it is observed that the GW response at the detector yields markedly different SNRs for the three scenarios. 
Without the presence of the dark matter halo, the dynamics of the system are essentially governed by the pure quasinormal oscillations.
The magnitude of the latter exhibits exponential decay followed by an inverse-power-law tail.
However, it is worth noting that this scenario does not correspond to the smallest characteristic strain.
With the presence of the dark matter halo, if the initial perturbation does not involve the matter field, the resulting characteristic strain stays primarily below that of the vacuum SMBH case.
This holds even though, in the time domain, the subsequent fluid mode overwhelms the ``would-have-been'' late-time tail in magnitude, as indicated by the red dashed curve and blue dash-dotted curves in Fig.~\ref{fig:40}.
This seemingly counterintuitive result is understood as the dark matter halo absorbing the energy carried by gravitational radiation, thereby suppressing the strength of the measured GW signals.
In particular, one observes that the characteristic strain of purely quasinormal oscillations sits slightly above the remaining scenarios for the region of relatively larger frequencies $f\ge 4\times 10^{-2}\text{Hz}$.
It is understood that the interaction with dark matter transfers a fraction of energy from the higher-frequency region of the spectrum associated with the black hole QNMs to lower frequencies related to the fluid mode, effectively suppressing characteristic strain in the former region.

The SNRs corresponding to the three different dark matter models show that steeper spikes result in lower SNRs. This is well understood: models with steeper spikes correspond to deeper effective potential wells, as can be seen from the expression of the effective potential. Propagating gravitational waves from a deeper potential well to infinity requires greater energy loss, which implies a weaker detectable signal strength. Considering the entire post‑merger waveform, including the $w$-mode and $f$-mode, even though a steeper dark matter model corresponds to a stronger $f$-mode, its overall amplitude is smaller.
Furthermore, in the context of the Taiji detector, the typical frequencies of the fluid modes and those of the BH QNMs both lie within the sensitive band of Taiji, and therefore both of them play a potential role in the detected GW signals, while QNM has a real part $\text{Re}\omega_0 \sim 4.77\times 10^{-2} \text{Hz}$, and the fluid mode, whose real part gives $\text{Re}\omega_f \sim 3.3\times 10^{-3} \text{Hz}$.

The above discussions are further reinforced by considering the scenario where the initial impact of a merger event on the dark matter distribution is manifested as the initial perturbations to the matter field, which leads to a potentially more pronounced effect on the emitted GWs.
As illustrated by the green dotted curve in Fig.~\ref{fig:40}, it overwhelms the other two scenarios up to $f\sim 1\times 10^{-2}\text{Hz}$.
This is attributed to the fact that the fluid modes are excited more significantly compared to the two other scenarios, particularly owing to the initial matter perturbations.

\begin{figure}[htbp] 
    \centering
    \includegraphics[width=0.8\textwidth]{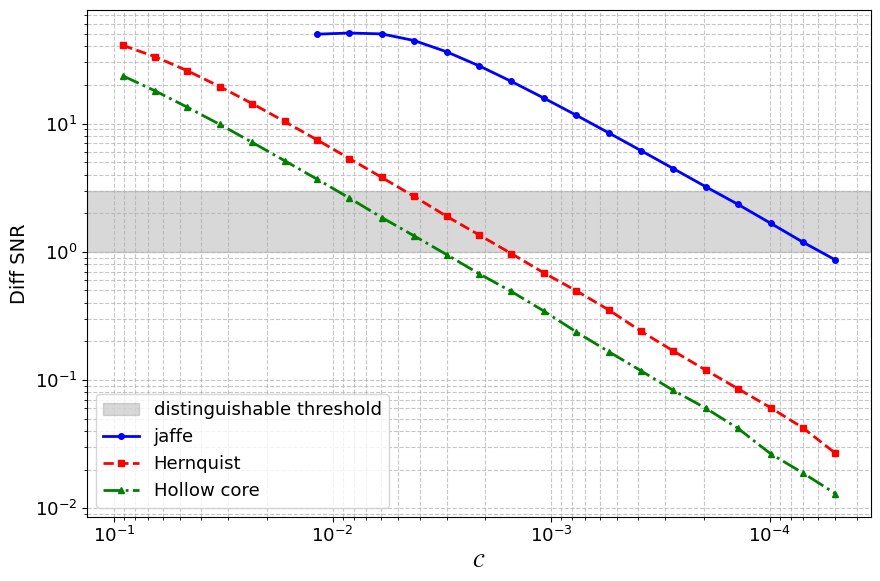}
    \caption{SNR of the difference waveform at a fixed signal amplitude for the three Dehnen-family profiles, relative to that of a vacuum SMBH. The results are shown as a function of the compactness $\mathcal{C}$, where $M_{{\mathrm{halo}}}=0.1{M}_{\mathrm{BH}}$.
    The gray band denotes the empirical threshold above which the waveform difference becomes distinguishable.}
    \label{fig:30}
\end{figure}

To this end, we evaluate the difference SNR to assess the relative signal strength with respect to that of a vacuum SMBH in Fig.~\ref{fig:30} for three different halo profiles. The difference SNR is defined as follows~\cite{lindblom2008model}:
\begin{equation}
\label{diff}
\rho_{\text{diff}}^{2} \equiv \,\big\langle h_1 -  h_2 \,\big|\, h_1 -  h_2 \big\rangle,
\end{equation}
where $h_1$ and $h_2$ are two GW waveform templates.
In the normalized and small-mismatch limit, $\text{Mismatch}\ll 1$, 
\begin{equation}
\rho_{\text{diff}} \approx \rho_{\text{opt}}\sqrt{2\,\text{Mismatch}} \, ,
\end{equation}
where $\rho_{\text{opt}}$ represents the optimal SNR of the event.

In this work we adopt $\rho_{\rm diff}\ge 1\,(3)$ as an empirical threshold for ``distinguishability'' (or ``significance'').
The results are presented as a function of compactness $\mathcal{C}$, defined as the ratio of the halo mass to its size
\begin{equation}\label{DefCompact}
\mathcal{C}=\frac{{M}_{\text{halo}}}{{a}_{0}}.
\end{equation}
Overall, the results indicate that for a given compactness, environment models with more pronounced central spikes have a stronger impact on the GW ringdown waveform. Moreover, this deviation from the vacuum waveform is nonlinear. As the spike steepens, the shift becomes increasingly pronounced. 
Among different profiles, it is shown that the Jaffe profile, which possesses a more pronounced spike, triggers the most significant difference in SNR Eq.~\eqref{diff}.
Notably, it clearly distinguishes itself from other profiles for compactness as low as $\mathcal{C}\simeq {10}^{-4}$, while existing studies suggest that realistic galactic halos have $\mathcal{C}\lesssim {10}^{-4}$~\cite{navarro1997universal,kim2004gravitational,spieksma2025black}. 
This implies that, as long as the underlying profile features a sufficiently sharp spike, it is promising to use the ringdown waveforms from typical events with SNRs between ${10}^{2}$ and ${10}^{3}$ to probe the properties of the dark matter halo.
For models with halo profiles of more moderate shape, the impact on the ringdown waveform is relatively suppressed, and the SNRs attainable with the currently planned GW detectors might be insufficient to extract quantitative information.
Furthermore, as mentioned, since the frequencies of both the fluid modes and QNMs reside in the detectable range of Taiji, they subsequently furnish sizable contributions in computing the difference SNR presented in Fig.~\ref{fig:30}.

\subsection{Bayesian inference for model parameters}\label{sec:3c}

Building on the analysis in the previous subsection and adopting a more quantitative perspective, we further employ statistical inference to estimate the parameters of the underlying gravitational system, including those of the dark matter profile, from the simulated GW signals.
In particular, we use matched filtering.
As a statistical inference approach, the Bayesian approach is not based on the frequentist point of view.
Instead, it aims to extract information on the probability of a hypothesis, typically expressed in terms of specific parameters of the underlying theoretical model, given observational data and prior knowledge.
Specifically, by applying Bayes' theorem, the analysis updates the posterior distribution of the model parameters of interest by combining the prior distribution with the likelihood of the observed data.
Formally, the posterior distribution is given by the formula:
\begin{equation}
P(\vec{\theta} \mid D) = \frac{\mathcal{L}(D \mid \vec{\theta})\,P(\vec{\theta})}{P(D)},
\end{equation}
where $P(\vec{\theta})$ is the prior distribution of a set of parameters in question, $\vec{\theta}$, based on prior knowledge, and $\mathcal{L}(D \mid \vec{\theta})$ is the likelihood function of the observational data $D$.
In the present study, one assumes that the uncertainties in the data are entirely attributed to the floor noise arising in the experimental measurements.
Subsequently, the difference between the data and theoretical value is essentially a random variable governed by the distribution of the floor noise, which can be reasonably approximated by a multi-dimensional Gaussian distribution~\cite{romano2017detection}. One can assume the observed data $\mathbf{d}$ consists of a deterministic signal model $\mathbf{h}(\vec{\theta})$ plus random noise $\mathbf{n}$:
\begin{equation}\mathbf{d} = \mathbf{h}(\vec{\theta}) + \mathbf{n}.\end{equation}
Then, the noise $\mathbf{n}$ follows a multivariate Gaussian distribution with a mean of zero and a covariance matrix $\mathbf{C}$. The probability density function for the noise is:
\begin{equation}
p(\mathbf{n}) = \frac{1}{\sqrt{\det(2\pi \mathbf{C})}} \exp\left( -\frac{1}{2} \mathbf{n}^T \mathbf{C}^{-1} \mathbf{n} \right).\end{equation}
The Likelihood $\mathcal{L}(D \mid \vec{\theta})$ is defined as the probability of observing the specific noise realization $\mathbf{n} = \mathbf{d} - \mathbf{h}(\vec{\theta})$.
So, in the frequency domain, the Likelihood $\mathcal{L}(D \mid \vec{\theta})$ can be written as:
\begin{equation}
\begin{aligned}
\ln \mathcal{L}(D \mid \vec{\theta}) = \sum_{f_{i}} \Bigg[ & -\frac{1}{2} (\tilde{\mathbf{d}} - \tilde{\mathbf{h}}(\vec{\theta}))^{T} \mathbf{C}^{-1} (\tilde{\mathbf{d}} - \tilde{\mathbf{h}}(\vec{\theta}))^{*} \\
& -\frac{1}{2} \ln(\det 2\pi \mathbf{C}) \Bigg],
\end{aligned}\label{likelihoodBaye}
\end{equation}
Where $\tilde{\mathbf{d}}(f_i)$ and $\tilde{\mathbf{h}}(\vec{\theta}; f_i)$ are column vectors of dimension $N_c=1024$ at frequency bin $(5\times 10^{-4},5\times 10^{-1})$. The superscripts $T$ and $*$ denote the transpose and complex conjugate operations, respectively.
Here, $\mathbf{C}(f_i)$ denotes the covariance matrix of the residual noises in the frequency domain:
\begin{equation}
\mathbf{C} = \frac{T_{\text{obs}}}{4}
\left[
\begin{array}{lll}
    S_{{AA}} & S_{{AE}} & S_{{AT}} \\
    S_{{EA}} & S_{{EE}} & S_{{ET}} \\
    S_{{TA}} & S_{{TE}} & S_{{TT}}
\end{array}
\right],
\end{equation}
where $T_{\text{obs}}$ represents the observation duration and $S_{12}$ denotes the cross-power spectral density between channels $1$ and $2$. 
In this study, since the $A$, $E$, and $T$ channels are constructed as orthogonal TDI combinations, the residual noise is assumed to be uncorrelated across channels. 
Consequently, the off-diagonal elements of $\mathbf{C}(f_i)$ are set to zero.
The summation in Eq.~\eqref{likelihoodBaye} is essentially an integral in the frequency domain, in the limit of a large sample size.

\begin{figure}[htbp] 
    \centering
    \includegraphics[width=0.8\textwidth]{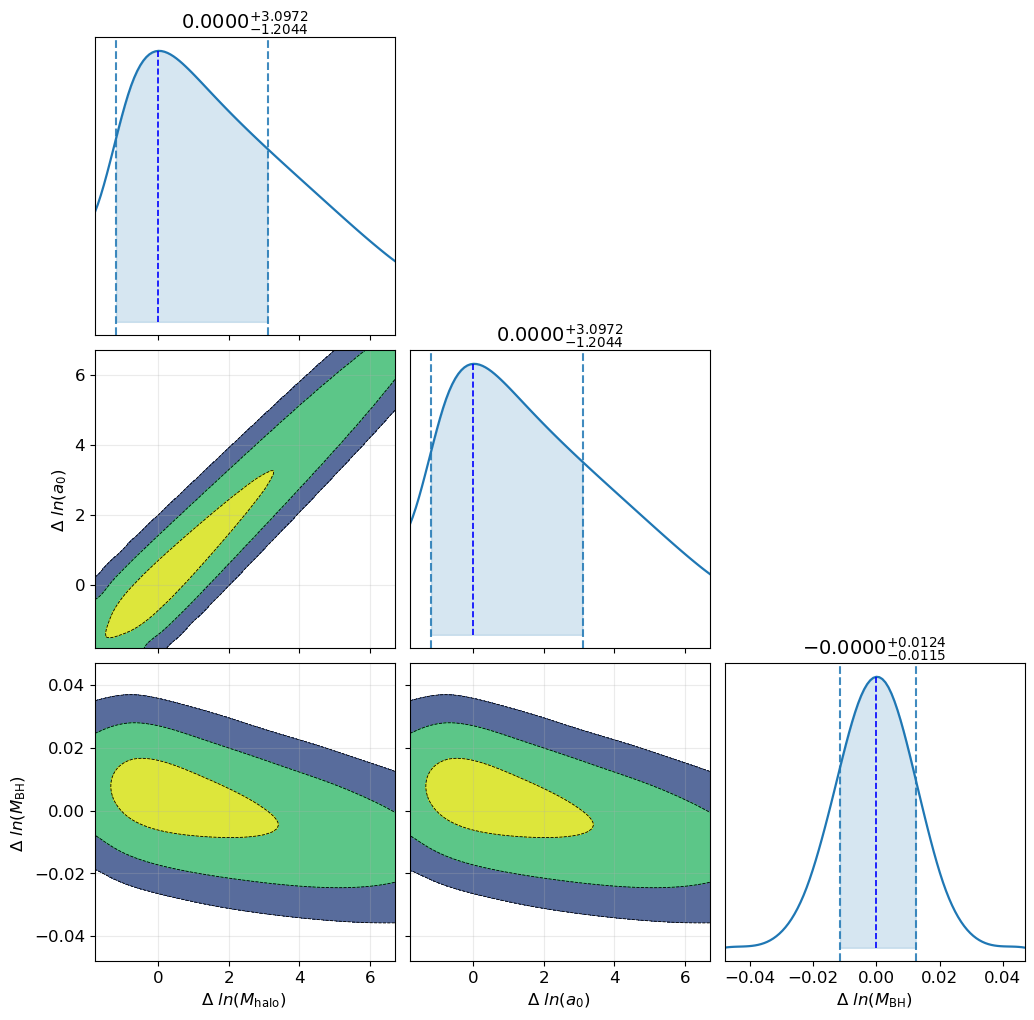}
    \caption{Corner plot of the posterior probability distributions for the parameters characterizing the dark-matter halo $ (M_\text{halo}, a_0) $ and the central black hole $ (M_\text{BH}) $.  
The GW signals are simulated using the Jaffe model with $ M_{\text{halo}} = M_{\text{BH}} $ and $ a_0 = 100 M_{\text{halo}} $.  
The three contour levels denote the $ 1\sigma $, $ 2\sigma $, and $ 3\sigma $ credible regions, while the diagonal panels display the one-dimensional marginals with shaded bands indicating the $ 68\% $ credible intervals.}
    \label{Fig:38}
\end{figure}

\begin{figure}[htbp] 
    \centering
    \includegraphics[width=0.8\textwidth]{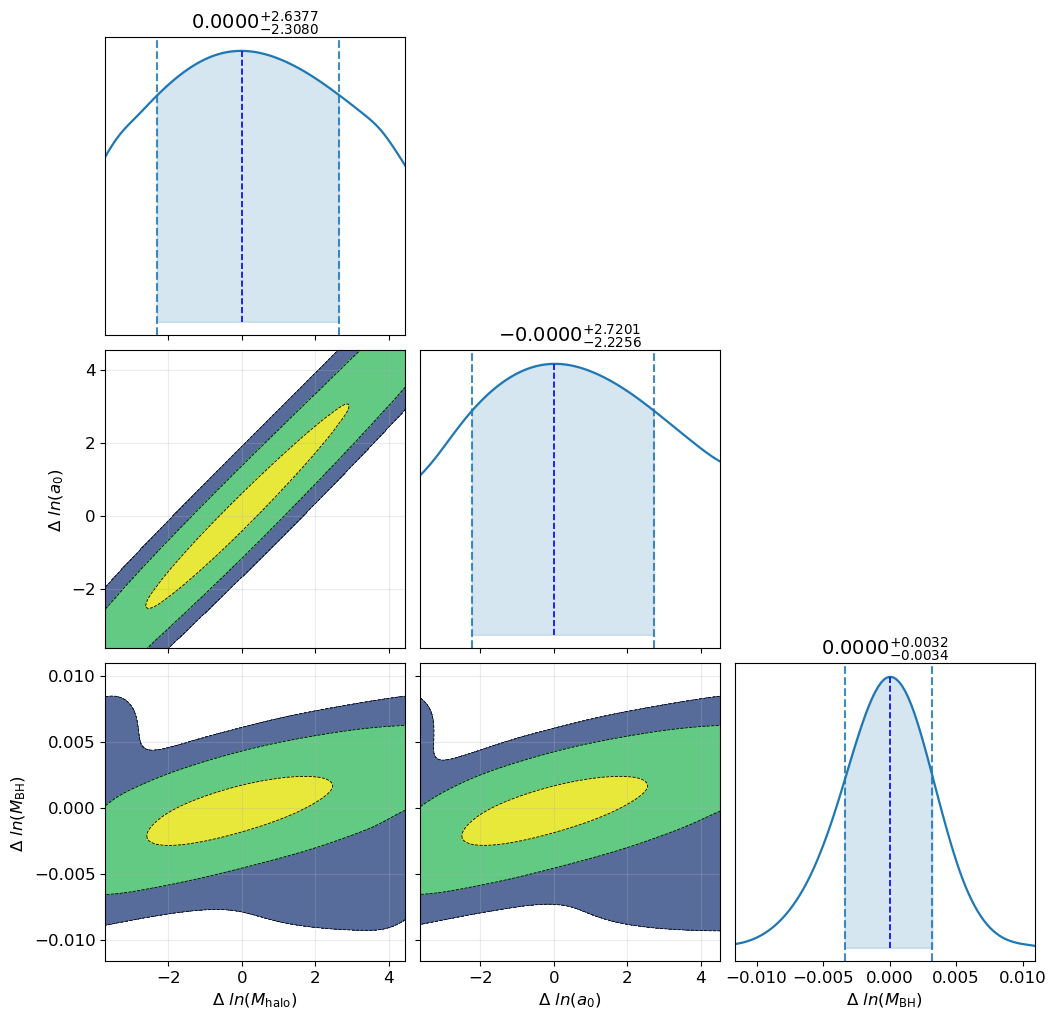}
    \caption{Corner plot of the posterior probability distributions for the parameters characterizing the dark-matter halo $ (M_\text{halo}, a_0) $ and the central black hole $ (M_\text{BH}) $.  
The GW signals are simulated using the Jaffe model with $ M_{\text{halo}} = M_{\text{BH}} $ and $ a_0 = 1000 M_{\text{halo}} $.  
The three contour levels denote the $ 1\sigma $, $ 2\sigma $, and $ 3\sigma $ credible regions, while the diagonal panels display the one-dimensional marginals with shaded bands indicating the $ 68\% $ credible intervals.}
    \label{Fig:45}
\end{figure}

\begin{figure}[htbp] 
    \centering
    \includegraphics[width=0.8\textwidth]{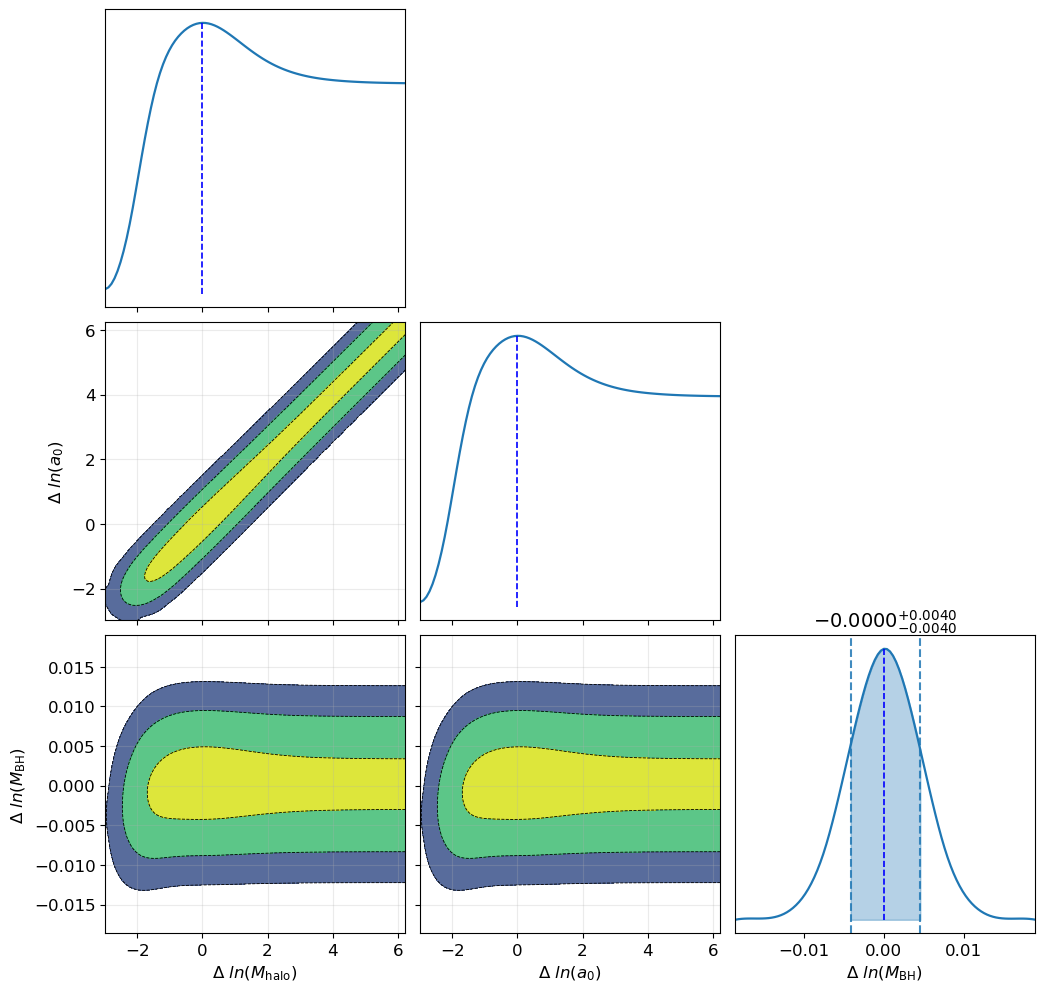}
    \caption{Corner plot of the posterior probability distributions for the parameters characterizing the dark-matter halo $ (M_\text{halo}, a_0) $ and the central black hole $ (M_\text{BH}) $.  
The GW signals are simulated using the Hernquist model with $ M_{\text{halo}} = 10M_{\text{BH}} $ and $ a_0 = 100 M_{\text{halo}} $.  
The three contour levels denote the $ 1\sigma $, $ 2\sigma $, and $ 3\sigma $ credible regions, while the diagonal is not possible to well define credible intervals for individual parameters due to parameter degeneracy.}
    \label{Fig:39}
\end{figure}

\begin{figure}[htbp] 
    \centering
    \includegraphics[width=0.8\textwidth]{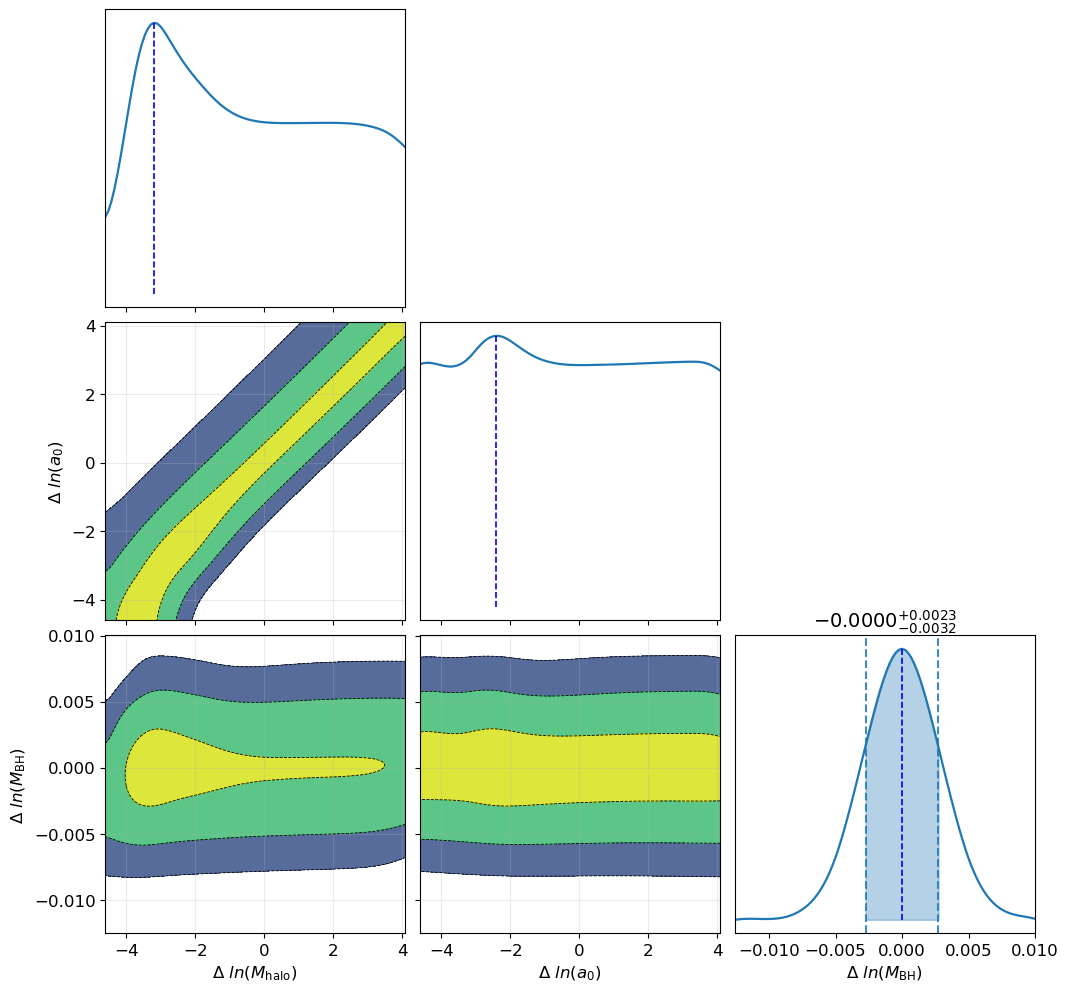}
    \caption{Corner plot of the posterior probability distributions for the parameters characterizing the dark-matter halo $ (M_\text{halo}, a_0) $ and the central black hole $ (M_\text{BH}) $.  
The GW signals are simulated using the Hernquist model with $ M_{\text{halo}} = 1M_{\text{BH}} $ and $ a_0 = 100 M_{\text{halo}} $.  
The three contour levels denote the $ 1\sigma $, $ 2\sigma $, and $ 3\sigma $ credible regions, while the diagonal is not possible to well define credible intervals for individual parameters due to parameter degeneracy.}
    \label{Fig:37}
\end{figure}

To infer the parameters of the underlying theoretical model, we will use the input data set, which comprises the data streams associated with the GW signals generated by the scenario without any initial density perturbations.
Here, we do not consider the scenario with an initial matter-field perturbation, as its modeling entails significant theoretical uncertainty.
By making use of the Bayesian inference via the likelihood given by Eq.~\eqref{likelihoodBaye}, we evaluate the posterior distribution and credible intervals for the parameters of the underlying black hole and dark matter halo, namely, the mass of the black hole $m_\text{BH}$, mass and size of the dark matter halo $m_\text{halo}$, and $a_0$.
The results are presented in Figs.~\ref{Fig:38}-\ref{Fig:37}. 

From the pairwise joint posterior distribution of the model parameters, we find that the two parameters $m_\text{halo}$ and $a_0$, which describe the properties of the dark-matter halo within the Dehnen family, are highly degenerate.  
In other words, while the joint posterior is sensitive to the compactness defined as the ratio between $m_\text{halo}$ and $a_0$ in Eq.~\eqref{DefCompact}, it becomes rather insensitive to either parameter individually once their ratio is fixed.  
Among different dark-matter profiles, the Jaffe model is observed to exhibit less parameter degeneracy than the Hernquist model.  
This implies that parameter inference is relatively more feasible for profiles possessing a more pronounced spike.  
In addition, comparing the two Hernquist profiles, it appears that the parameter degeneracy tends to be less severe for dark-matter halos with higher compactness.  
On the other hand, for the mass of the central SMBH, one obtains a well-constrained posterior distribution.  
Specifically, for the same true value of dark matter profiles, one has $\sigma_{{\rm ln} {M}_{\text{BH}}} \approx 3.0\times 10^{-3}$ of the Hernquist model, whereas for the Jaffe model the uncertainty increases to $1.9\times 10^{-2}$, nearly an order of magnitude larger.  
This result may be interpreted as indicating that, when the central SMBH is tightly enveloped by a pronounced spike with a comparable mass located near its horizon, the estimation of the black hole mass becomes less precise.  

\begin{figure}[htbp]
    \centering
    \begin{subfigure}[b]{0.48\textwidth} 
        \includegraphics[width=\linewidth]{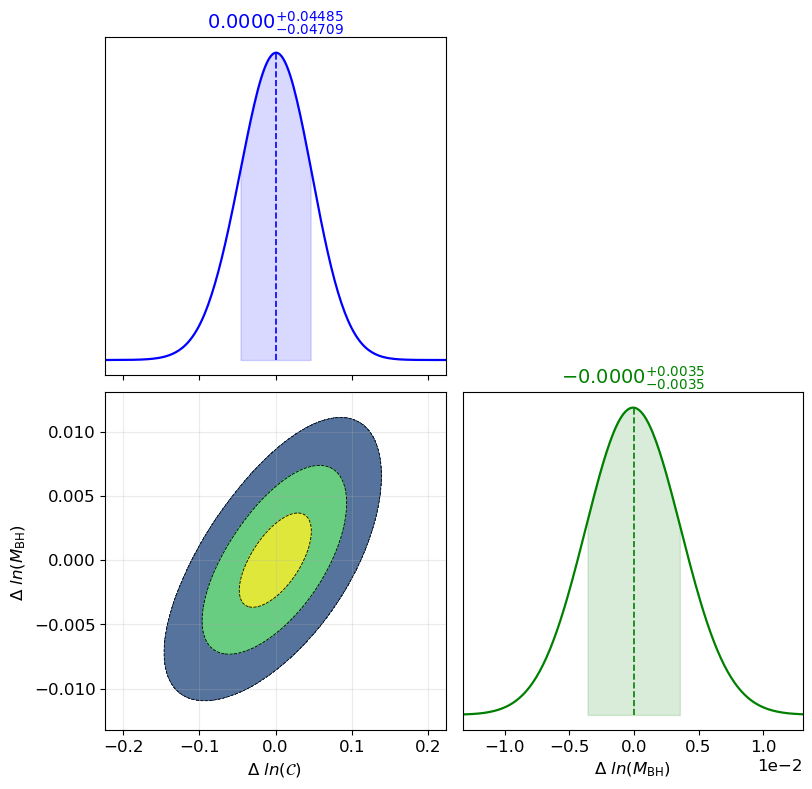} 
        \caption{Hernquist: $M_{\text{halo}}=10M_{\text{BH}},a_0=100M_{\text{halo}}$}
        \label{fig:32}
    \end{subfigure}
    \hfill 
    \begin{subfigure}[b]{0.48\textwidth}
        \includegraphics[width=\linewidth]{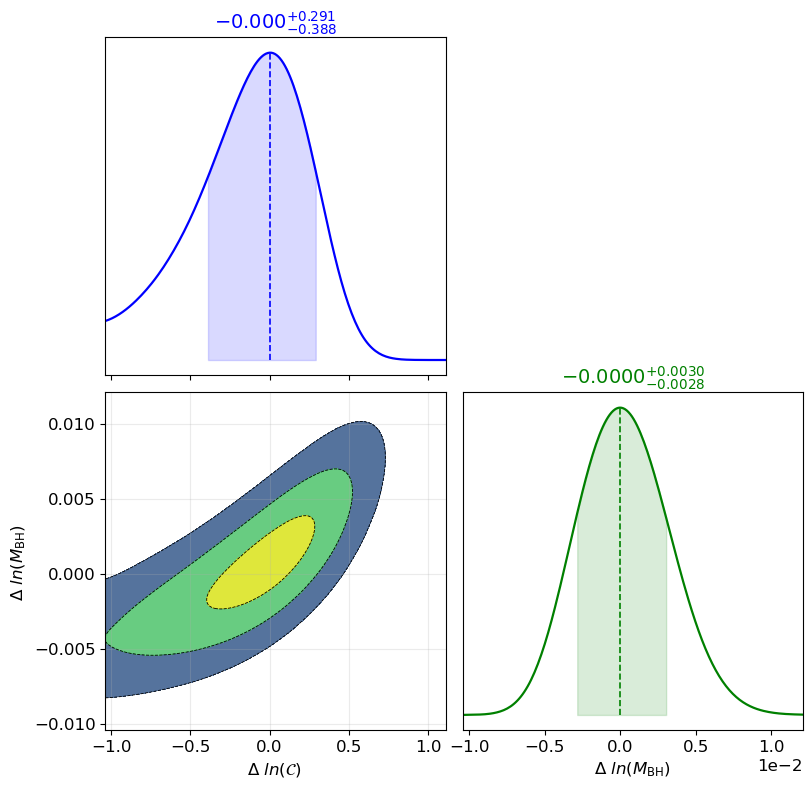}
        \caption{Hernquist: $M_{\text{halo}}=1M_{\text{BH}},a_0=100M_{\text{halo}}$}
        \label{fig:34}
    \end{subfigure}
    \begin{subfigure}[b]{0.48\textwidth}
        \includegraphics[width=\linewidth]{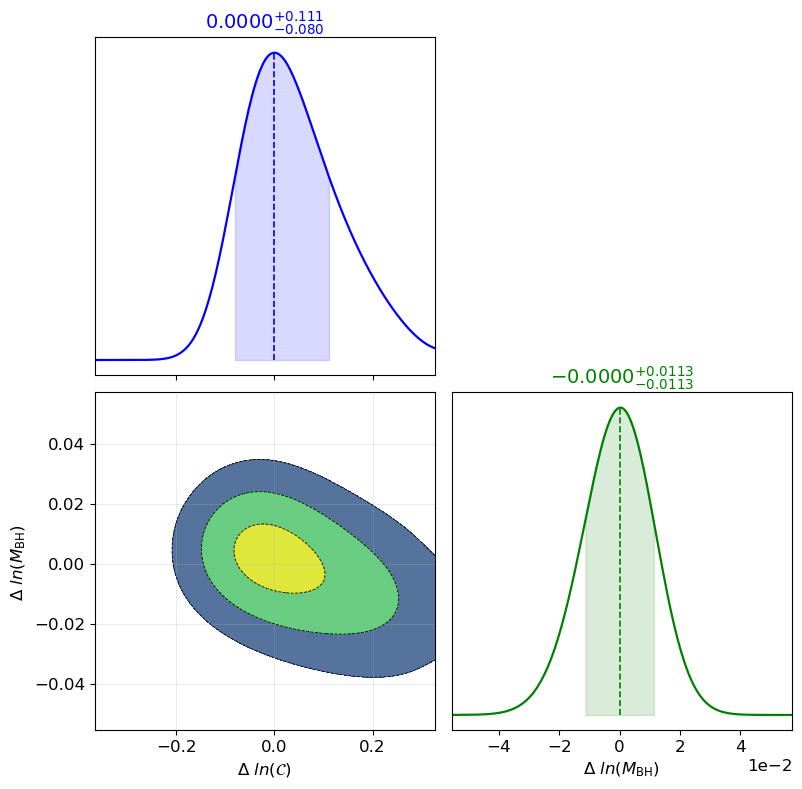}
        \caption{Jaffe: $M_{\text{halo}}=1M_{\text{BH}},a_0=100M_{\text{halo}}$}
        \label{fig:33}
    \end{subfigure}
    \begin{subfigure}[b]{0.48\textwidth}
        \includegraphics[width=\linewidth]{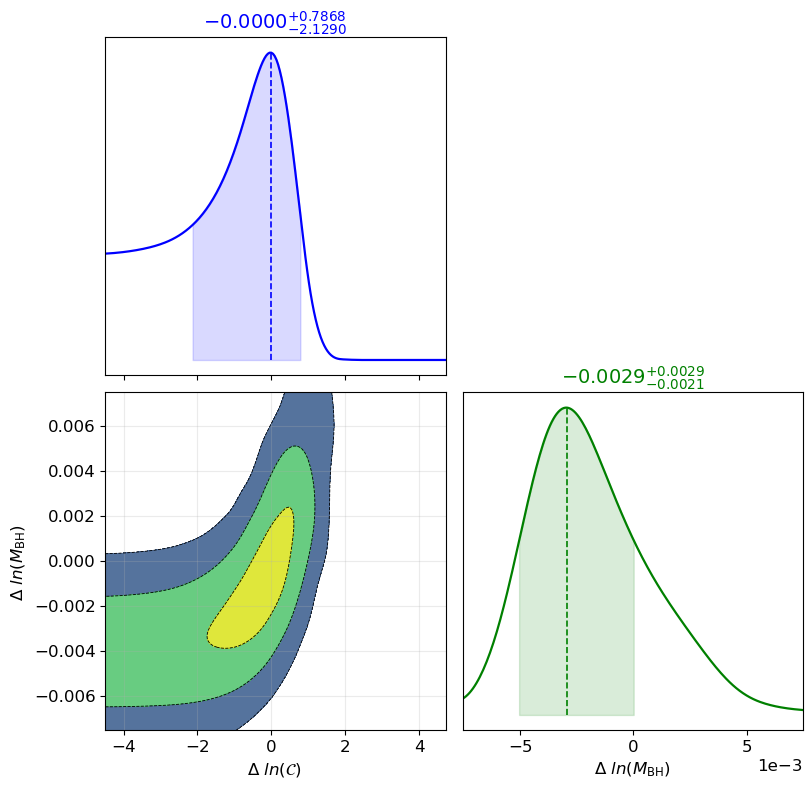}
        \caption{Hollow core: $M_{\text{halo}}=1M_{\text{BH}},a_0=100M_{\text{halo}}$}
        \label{fig:35}
    \end{subfigure}
    \caption{Corner plot of the posterior probability distributions for the parameters characterizing the dark-matter halo $ \mathcal{C} $ and the central black hole $ (M_\text{BH}) $.  
The GW signals are simulated using three different models and different true values.  
The three contour levels denote the $ 1\sigma $, $ 2\sigma $, and $ 3\sigma $ credible regions, while the diagonal panels display the one-dimensional marginals with shaded bands indicating the $ 68\% $ credible intervals.
} 
    \label{twoD}
\end{figure}

Regarding the Bayesian inference, from a practical point of view, the degeneracy revealed above among the parameters characterizing the dark-matter halo indicates that it may be more efficient to focus on the dimensionless compactness parameter and reserve computational resources for the remaining features of interest.  
In Fig.~\ref{twoD}, we illustrate the posterior estimates for the compactness and black hole mass, $\mathcal{C}$ and ${M}_{\text{BH}}$, and examine more closely the differences among various halo profiles, with additional results summarized in Table~\ref{sigma}.  
Across all models, we confirm our previous observation that the sharper the spike, the more accurately the compactness is recovered, while the precision of the estimated ${M}_{\text{BH}}$ is progressively degraded.  
Moreover, as $\mathcal{C}$ decreases, its uncertainty $\sigma_{ \ln \mathcal{C}}$ increases, whereas that of the black hole mass, $\sigma_{ \ln M_{\text{BH}}}$, decreases.  
In fact, the uncertainties of the inferred $\mathcal{C}$ and ${M}_{\text{BH}}$ depend primarily on $\mathcal{C}$ itself, and only weakly on the individual halo parameters.  

\begin{table}[ht]
\centering
\begin{tabular}{|c|c|c|c|c|c|c|}
\hline
\multirow{2}{*}{($M_{\text{halo}}$, $a_0$)} & \multicolumn{3}{c|}{$\sigma _{ ln\mathcal{C}}$} & \multicolumn{3}{c|}{$\sigma _{ln M_{\text{BH}}}$} \\ \cline{2-7} 
                                      & Jaffe & Hernquist & Hollow core & Jaffe   & Hernquist & Hollow core
                                      \\ \hline
(1, $10^2$)                           & $9.5\times{ 10}^{-2}$      & $3.4 \times 10^{-1}$  & $1.5 $  & $1.1\times{ 10}^{-2}$     & $2.9 \times 10^{-3}$  & $2.5 \times 10^{-3}$  \\ \hline
(1, $10^3$)                          & $1.4 \times 10^{-1}$      & -  & -  & $3.2 \times 10^{-3}$      & -  & -  \\ \hline
(10, $10^2$)                         & -      & $2.5 \times 10^{-2}$  & $7.3 \times 10^{-2}$  & -      & $3.5 \times 10^{-3}$  & $3.2 \times 10^{-3}$  \\ \hline
(10, $10^3$)                         & $1.0 \times 10^{-1}$      & $3.3 \times 10^{-1}$   & 1.1  & $1.1 \times 10^{-2}$      & $3.0 \times 10^{-3}$  & $2.6 \times 10^{-3}$  \\ \hline
\end{tabular}
\caption{Estimation errors on \(\ln\mathcal{C}\) and \(\ln M_{\text{BH}}\) for the three Dehnen-family dark-matter models under different true values.  
An em-dash “-” indicates that no reliable error could be obtained, either because the numerical results diverged or because the waveform became indistinguishable from the vacuum case.}
\label{sigma}
\end{table}

Moreover, for the Jaffe model, even when the compactness \( \mathcal{C} \) is on the order of \( 10^{-3} \), the estimation error for \( \ln \mathcal{C} \) is only \( 1.4\times 10^{-1} \) (which translates back to \( 1.5\times 10^{-4} \) for \( \mathcal{C} \) itself). Therefore, even with a tiny \( \mathcal{C} \), extreme spike models can use our method for a rapid and accurate posterior parameter estimation. Moreover, this method is efficient because the ringdown phase following the merger is brief, and even when accounting for the fluid modes, the total data duration amounts to only about an hour. Being able to quickly determine the compactness \( \mathcal{C} \) of the dark-matter halo surrounding a merger event using this data would be a very cost-effective and meaningful task for subsequent observations and data processing.

\section{Further discussions and concluding remarks}\label{sec:4}

Although spectral instability leads to a significant deformation in the QNM spectrum in the frequency domain, it is understood that its impact on the time-domain waveform is, by and large, limited (see, in particular, recent analysis by Wu {\it et al.}~\cite{agr-qnm-instability-83} which is a qualitative refinement of previous findings~\cite{agr-qnm-instability-02, agr-qnm-instability-11}).
This also leads to the discussion about the physical significance of metric perturbations that give rise to such spectral instability~\cite{agr-qnm-instability-16}.
In this context, the emergence of the fluid modes in the time-domain ringdown waveform is a pertinent topic with observational implications.
In this work, we analyze the feasibility of inferring the properties of the gravitational system formed by an SMBH merging with a dark matter halo via the emitted GW signals.  
On the theoretical side, the study relies on a fully relativistic approach recently established by Cardoso {\it et al.}~\cite{cardoso2022gravitational}, in which the gravitational sector effectively interacts with the matter field through the minimal coupling in the action.  
Moreover, we extend the existing metric to incorporate a class of Dehnen-type halo profiles.  
By considering three distinct scenarios, we simulate GWs by numerically solving the system of dynamical equations with initial perturbations imposed on both the metric and the matter field.  
The metric perturbations evaluated in the Regge–Wheeler gauge are transformed into the TT gauge, where they become asymptotically plane waves with well-defined polarization states, providing an appropriate description of the spacetime deformations captured by a GW detector. 
On the experimental side, we explicitly implemented realistic aspects such as instrumental noises and TDI algorithms.
Specifically, a few typical second-generation TDI combinations are considered so that the embedded GW signals, used as the input dataset for the Bayesian inference, are properly time-shifted and superposed, offering a more realistic data-processing setup.  
In this study, the data analysis is carried out using specific parameter sets for the Taiji detector, but it can be straightforwardly generalized to other ongoing space-borne GW detection programs.  
Given the ongoing efforts on both the theoretical and experimental fronts, the present study offers a meaningful assessment of the feasibility of this approach.  

The main features of the gravitational system explored in the present work are twofold.  
Firstly, it relies on an analytic solution in which the polar gravitational perturbation interacts with the matter field as a result of minimal coupling, so that the resulting GW signals naturally encode essential information about the matter distribution.  
A distinctive attribute of the latter is the emergence of fluid modes, which persist far longer in the time domain and exhibit stronger amplitudes.  
Numerically, such modes appear near the lower end of the GW detector’s relevant frequency window.  
The presence of the dark matter halo induces a transition from the initial quasinormal oscillations to the fluid modes, overtaking the late-time tail characteristic of a pure SMBH.  
In the frequency domain, this transition effectively produces a shift toward lower frequencies in the characteristic strain and SNR, leading to a nontrivial impact on the signal that could be detected by future space-borne GW observatories.  
Secondly, although it lies beyond the scope of the present study, EMRIs have been extensively studied in this context in the literature.  
When comparing these two physical observables, the present work focuses primarily on ringdown signals emitted on much shorter timescales than those of EMRIs, whose observations typically span long periods.  

Bayesian inference is performed using the simulated GW signals as the input dataset.  
Our analysis reveals that the spike feature in the dark matter profile leaves a distinctive imprint on the resulting GW signals, which can be identified through the inference procedure.  
Although some degeneracy arises among the parameters governing the specific form of the dark matter profile, our numerical results remain encouraging for extracting meaningful information about the underlying gravitational system that may have undergone a merger process.  
We show that key parameters, such as the mass and compactness, can be reliably distinguished by means of Bayesian analysis.  
These results further indicate that the presence of dark matter is likely to enhance the detectability of such signals.  
We plan to pursue further developments along this line in future studies.

\section{acknowledgment}

This work is supported by the National Key Research, Development Program of China, Grant No. 2021YFC2201901, 2021YFC2201903, and the International Partnership Program of the Chinese Academy of Sciences, Grant No. 025GJHZ2023106GC.
We also acknowledge the financial support from Brazilian agencies 
Funda\c{c}\~ao de Amparo \`a Pesquisa do Estado de S\~ao Paulo (FAPESP), 
Funda\c{c}\~ao de Amparo \`a Pesquisa do Estado do Rio de Janeiro (FAPERJ), 
Conselho Nacional de Desenvolvimento Cient\'{\i}fico e Tecnol\'ogico (CNPq), 
and Coordena\c{c}\~ao de Aperfei\c{c}oamento de Pessoal de N\'ivel Superior (CAPES). 
For M.T., this research was funded by the Polish National Science Center Grant No.  2023/49/B/ST9/02777. M.T. thanks the National Institute for Space Research (INPE, Brazil) for their kind hospitality while this work was done.

\bibliography{references_liang, references_qian}

\end{document}